\documentclass[aps, prl, reprint, amsmath, amssymb, tbtags]{revtex4-2}

\usepackage{siunitx}
\usepackage{bm}
\usepackage{graphicx}
\usepackage[utf8]{inputenc}
\usepackage{comment} 
\usepackage{mathtools}
\usepackage{braket}

\usepackage{soul}
\usepackage{subcaption}

\usepackage[colorlinks=true,
            linkcolor=blue,
            citecolor=blue,
            urlcolor=blue]{hyperref}

\renewcommand{\vec}{\bm} 
\renewcommand{\Im}[1]{\text{Im}\{#1 \}}
\renewcommand{\Re}[1]{\text{Re}\{#1 \}}
\renewcommand{\d}{\mathrm{d}}
\newcommand{\iu}{\mathrm{i}}
\newcommand{\e}{\text{e}}
\newcommand{\dotvec}[1]{\dot{\vec #1}}

\newcommand{\vrho}{\boldsymbol \rho} 
\newcommand{\mo}{\tilde{\vec m}_0}
\usepackage{soul}

\usepackage{xcolor}

\definecolor{Red}{rgb}{1,0,0}

\begin{document}
\title{Nutational Spin Pumping and Dissipation}
\author{Hans Gløckner Giil}
\author{Arne Brataas}
\affiliation{Center for Quantum Spintronics, Department of Physics, Norwegian University of Science and Technology, NO-7491 Trondheim, Norway}
\date{\today}
\begin{abstract}
At terahertz drive frequencies, magnetization dynamics enter an inertial regime beyond the scope of standard Landau–Lifshitz–Gilbert (LLG) theory.
We reveal a nutational damping torque comprising a bulk term and an interfacial contribution that describes spin emission into adjacent metals.
The interfacial term generates nutational spin pumping: a dc spin current that scales quadratically with the drive frequency.
Using a scattering approach, we relate this effect to microscopic material parameters and predict pronounced enhancements in ultrathin films and low–Fermi-energy conductors.
Nutational spin pumping can be detected directly via the inverse spin Hall effect or indirectly via thickness-dependent renormalization of nutational damping.
\end{abstract}
\maketitle
\textit{Introduction---} 
The Landau-Lifshitz-Gilbert (LLG) equation has long served as the cornerstone for understanding magnetization dynamics in ferromagnets (FMs), accurately describing precessional motion and damping at \si{\giga \hertz} frequencies \cite{landauTheoryDispersionMagnetic1935, gilbertClassicsMagneticsPhenomenological2004}. 
Its success is supported by extensive experimental and theoretical evidence that demonstrates its robustness across a wide range of materials and device architectures. However, the push toward ultrafast spintronics has brought increasing attention to the \si{\tera \hertz} regime, where traditional descriptions fail to capture essential high-frequency physics. 

Recent theoretical advances have revealed that, at these high frequencies, magnetization acquires an effective inertia, modifying the LLG equation to include second-order time derivatives, capturing an inertial effect that mirrors the behavior of mass in mechanical systems and results in nutational motion in addition to precession~\cite{ciorneiMagnetizationDynamicsInertial2011, mondalInertialEffectsUltrafast2023}.
This inertial response arises microscopically from various mechanisms, including spin–orbit coupling, electronic spin dynamics, surface-induced anisotropy, and exchange processes in heterostructures \cite{fahnleGeneralizedGilbertEquation2011, wegroweMagnetizationDynamicsGyromagnetic2012, mondalGeneralisationGilbertDamping2018, quarentaBathInducedSpinInertia2024, bhattacharjeeAtomisticSpinDynamic2012a, kikuchiSpinDynamicsInertia2015, johnsenDynamicallyGeneratedSpin2025, bastardisMagnetizationNutationInduced2018a}.
Recent studies develop spin inertia for anisotropic ferromagnets \cite{cherkasskiiTheoryInertialSpin2022} and introduce a tensorial generalization beyond the scalar description \cite{ghoshTheoryTensorialMagnetic2024}.
Other advances include spin switching due to exchange-induced inertia in antiferromagnets (AFMs)~\cite{kimelInertiadrivenSpinSwitching2009}, inertial spin waves and their resonance frequencies~\cite{mondalInertialSpinWaves2022, cherkasskiiDispersionRelationNutation2021, cherkasskiiNutationResonanceFerromagnets2020, lomonosovAnatomyInertialMagnons2021, titovNutationSpinWaves2022}, the proposal of auto-oscillations in inertial FMs~\cite{rodriguezSpinInertiaAutoOscillations2024}, and nutational switching \cite{winterNutationalSwitchingFerromagnets2022} for ultrahigh-frequency technology applications.
Finally, investigations of conventional spin pumping at nutational resonances, exhibiting the familiar $\propto \omega$ scaling~\cite{mondalSpinPumpingTerahertz2021}, have also provided useful context for the bulk and interfacial nutational dissipation addressed in this Letter.

Several experimental studies have established inertia‑driven nutation in ferromagnets~\cite{neerajInertialSpinDynamics2021, unikandanunniInertialSpinDynamics2022, deMagneticNutationTransient2025, liInertialTermsMagnetization2015}, highlighting the growing importance of spin inertia in ultrafast magnetization dynamics. Although these works have clarified the conservative aspects of inertial motion, the corresponding \textit{dissipative} phenomena that arise at THz frequencies remain largely unexplored. This lack of understanding of high‑frequency dissipative dynamics is the central motivation for our study.
\begin{figure}
    \centering
    \includegraphics[width=0.8\linewidth]{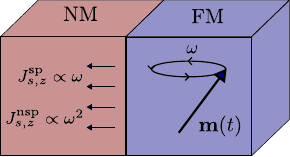}
    \caption{
    The precessing magnetization pumps spin currents into the normal metal with two contributions, the conventional contribution $J_{s, z}^{\text{sp}}$ linear in the driving frequency $\omega$ and the nutational contribution $J_{s, z}^{\text{nsp}}$ quadratic in $\omega$.
    Both contributions exhibit resonance peaks for right-handed driving fields at the FMR frequency, and for left-handed driving fields at the NR frequency.
    }
    \label{fig:1}
\end{figure}

When magnetic materials, whether insulating, semiconducting, or metallic, are placed in contact with conductors, their dynamic magnetization can inject spin currents into adjacent layers, a phenomenon known as spin pumping~\cite{tserkovnyakEnhancedGilbertDamping2002, tserkovnyakSpinPumpingMagnetization2002, tserkovnyakNonlocalMagnetizationDynamics2005}. 
This effect plays a central role in modern spintronics by enabling the generation of pure spin currents without charge flow. Spin-pumped currents not only form the foundation of energy-efficient spin-based circuits but also serve as sensitive probes of the underlying spin dynamics. 
Critically, this spin emission leads to nonlocal dissipation: the loss of angular momentum to the environment, which modifies the magnet's damping. 
Recent studies show that spin pumping can even renormalize the gyromagnetic ratio, signaling deeper consequences beyond simple energy loss~\cite{khodadadiEnhancedSpinPumping2017}. 
This reveals the dual nature of spin pumping: it comprises both conservative components, which are time-reversal-invariant, and dissipative contributions, which are time-reversal-odd, underscoring its profound role in magnetization dynamics and energy transfer across interfaces.

With mounting evidence that spin inertia dominates magnetization dynamics at THz frequencies, and that spin pumping at GHz frequencies significantly modifies both reactive and dissipative torques, a natural question arises: Does a novel form of nutational damping and spin pumping emerge in the ultrafast THz regime? In this Letter, we demonstrate that it does.

In this Letter, we widen the understanding of ultrafast magnetization dynamics in three ways.
First, we extend the conventional inertial LLG equation to include a nutational dissipation term, an analog to Gilbert damping in higher-order dynamics, and demonstrate that it modifies the linewidth of nutational spin modes while leaving the resonance frequencies largely unaffected.
The nutational dissipation term has bulk and interface contributions, analogous to spin-pumping-enhanced Gilbert damping~\cite{tserkovnyakEnhancedGilbertDamping2002}.
Second, we uncover interface-driven nutational spin-pumping mechanisms that directly contribute to spin inertia dissipation and propose experimental probes to measure this effect, either directly via the pumped spin current or indirectly via the enhanced nutational damping in the FM.
Third, we derive microscopic expressions for these contributions, yielding quantitative estimates of their magnitudes. 
We compute the nutational-pumped spin current as a second-order contribution in the drive frequency, thereby going beyond the frozen-scattering-matrix formulation, as in e.g., Ref.~\onlinecite{brouwerScatteringApproachParametric1998}. 
These results reveal an ultrafast interfacial spin‑transport regime in which higher‑order dynamics govern angular‑momentum transfer.

\textit{Inertial Landau-Lifshitz-Gilbert equation}--- 
The most general norm-preserving form of the inertial LLG equation, to second order in temporal variations, is given by
\begin{align} \nonumber
    \dotvec m =& - \gamma \mu_0  \vec m \times  \vec H + \alpha \vec m \times  \dotvec m 
    + \eta \vec m \times   \ddot{\vec m}
    \\ \label{eq:ILLG}
    &
    + \alpha_\eta \eta  \vec m \times  (\vec m \times \ddot{\vec m}),
\end{align}
where $\gamma > 0 $ is the gyromagnetic ratio, $\vec H$ is the effective field, and $\alpha$ is the Gilbert damping. The third term on the right-hand side of Eq.~\eqref{eq:ILLG} describes the standard, reactive (invariant under time reversal) inertia dynamics. $\eta$ is an inertial relaxation time; $\omega_\eta = 1/ \eta$ is the frequency scale of the inertial regime \cite{ciorneiMagnetizationDynamicsInertial2011, fahnleGeneralizedGilbertEquation2011, mondalInertialEffectsUltrafast2023, thonigMagneticMomentInertia2017, oliveFerromagneticResonanceInertial2012, oliveDeviationLandauLifshitzGilbertEquation2015}. 
Our first main result is the term proportional to the dimensionless constant $\alpha_\eta$, which is dissipative (odd under time reversal) and not present in the usual discussions of spin inertia. 
To our knowledge, this dissipative term has neither been proposed nor analyzed previously.
We note that the inertial damping term is distinct from nonlinear damping corrections to the Gilbert torque~\cite{tiberkevichNonlinearPhenomenologicalModel2007}, which arise from amplitude-dependent effects beyond linear response, whereas the present contribution appears already at the linear level as a higher-order temporal derivative.
In the Supplemental Material (SM)~\cite{SeeSupplementalMaterial}, we derive Eq.~\eqref{eq:ILLG} using a phenomenological approach.
Eq.~\eqref{eq:ILLG} is a minimal low-energy effective theory: it retains the two lowest temporal derivatives of the magnetization, which is sufficient to capture the main low-energy phenomena, namely the FMR and nutational modes together with their dissipation. 
Here, $\alpha$ and $\alpha_\eta$ are independent symmetry-allowed coefficients rather than successive terms in a frequency expansion of a single damping function $\alpha(\eta\omega)$.

We will demonstrate that in the linearized regime, Eq.~\eqref{eq:ILLG} yields a low‑frequency mode described by the resonance frequency $\omega_H$ and damping governed by $\alpha$, and a high‑frequency nutation described by a resonance frequency $\omega_\eta$ with damping given by $\alpha + \alpha_\eta$.
$\alpha_\eta$ has a small effect on the nutational frequency but contributes significantly to the nutational damping, making it qualitatively distinct from the Gilbert damping that governs precessional motion. 
Crucially, it is the term proportional to $\alpha_\eta$ that enables nutational spin pumping, our second main result, which is quadratic in drive frequency. This nutational spin pumping enhances the nutational damping parameter $\alpha_\eta$, similar to how standard spin pumping enhances the Gilbert damping parameter $\alpha$ \cite{tserkovnyakNonlocalMagnetizationDynamics2005}.

\textit{Ferromagnetic and nutational resonance}---
We obtain the ferromagnetic resonance (FMR) {\it and} nutational resonance (NR) frequencies of Eq.~\eqref{eq:ILLG} by applying a polarized external field $\vec H(t) = (h_x(t), h_y(t), H_z)$, where $H_z > 0$ is an external field, constant in time. 
In the linearized regime (regarding $h_x$ and $h_y$ as small quantities), $\vec m \simeq \vec z + \delta \vec m$, with $\delta \vec m$ as a two-component vector in the $xy$-plane. Eq.~\eqref{eq:ILLG} can then be written compactly on a circular basis with $\delta m_\pm = \delta m_x \pm \iu \delta m_y$ and $h_\pm = h_x \pm \iu h_y$:
\begin{align} 
    \left[\eta ( 1 \pm \iu \alpha_\eta) \partial_t^2 + ( \alpha \pm \iu) \partial_t +  \omega_H \right] \delta m_\pm 
    \label{eq:llg_circbas}
    =  \gamma \mu_0 h_\pm,
\end{align}
where we introduced $\omega_H = \gamma \mu_0 H_z$ for notational convenience.

Because our interest lies in the susceptibility describing the steady‑state magnetization response, we Fourier-transform Eq.~\eqref{eq:llg_circbas}. 
In frequency space ($\omega$), the time derivatives become algebraic factors, allowing us to solve for $\delta m_\pm(\omega) = \chi_\pm( \omega) h_\pm(\omega)$. 
The dynamic susceptibilities are $\chi_\pm(\omega) = (\gamma \mu_0) [- \omega^2 \eta ( 1 \pm \iu  \alpha_\eta) + \iu \omega ( \alpha \pm  \iu ) +  \omega_H]^{-1}$.
The eigenvalues of the susceptibility matrix yield the magnetization resonance frequencies, including both the dissipative and reactive effects of the ferromagnetic resonance and inertia-driven nutation.
The susceptibilities $\chi_\pm$ have low-frequency modes dominated by $\omega_H$ and $\alpha$, and high-frequency modes governed by $\eta$ and $\alpha_\eta$. 
Here, we reproduce the form of the precessional and nutational resonances in the limit where the frequencies are well separated, allowing us to solve for the high-frequency modes by disregarding the $\omega$-independent term and the low-frequency modes by disregarding the $\omega^2$ term:
\begin{subequations}
\begin{align}
    \omega_\pm^\text{prec} & = \frac{\omega_H}{1 + \alpha^2} [\pm 1    
    + \iu \alpha],
    \\ 
    \omega_\pm^\text{nut} &= \frac{\omega_\eta}{1 + \alpha_\eta^2}
    \left[\mp
     \left(1 - \alpha \alpha_\eta \right)
    + \iu  (\alpha + \alpha_\eta)\right].
\end{align}
\label{eq:resonances}
\end{subequations}
We identify $\omega^\text{prec}$ with the conventional low-frequency precessional FMR branch, and $\omega^\text{nut}$ with a distinct nutational eigenmode at the much higher frequency $\omega_\eta$. 
The upper (lower) sign in Eq.\eqref{eq:resonances} corresponds to opposite chiralities of the driving field $h_\pm$, and shows that the chirality of the nutational modes is opposite to that of the precessional mode: a right-handed drive can excite a positive-frequency FMR mode, while a left-handed drive can excite a positive-frequency NR mode.

A central and previously unrecognized result of our theory is that the inertial damping parameter $\alpha_\eta$ selectively renormalizes the dissipation of the nutational mode while leaving the FMR damping unchanged. In the limit $\alpha_\eta=0$, our equations reduce to the familiar two-branch spectrum and recover the standard low- and high-frequency solutions reported in Ref.~\onlinecite{mondalSpinPumpingTerahertz2021}. 
Importantly, the experimentally observed linewidth hierarchy reported in Refs.~\onlinecite{neerajInertialSpinDynamics2021,unikandanunniInertialSpinDynamics2022}, namely that the nutational mode exhibits a substantially broader linewidth than the FMR mode, provides direct support for our prediction and is naturally explained by our eigenfrequencies and linewidths in Eq.~\eqref{eq:resonances}: the nutational linewidth is controlled by $\alpha + \alpha_\eta$, whereas the FMR linewidth is governed solely by $\alpha$.
In Ref.~\cite{unikandanunniInertialSpinDynamics2022}, the nutational linewidth is reported to be approximately one order of magnitude larger than that of the precessional mode.
Within our framework, this observation is naturally accounted for by a value of $\alpha_\eta$ that exceeds $\alpha$ by a similar factor.

From a phenomenological perspective, we expect the inertial terms to comprise both bulk and spin-pumping contributions.
This is analogous to conventional damping decomposition in ferromagnet–normal metal systems, where the total Gilbert damping coefficient $\alpha = \alpha^{\text{b}} + \alpha^\text{sp}$ has bulk contributions ($\alpha^{\text{b}}$) and a nonlocal spin-pumping-enhanced damping contribution ($\alpha^\text{sp}$). 
Analogously, we posit that there are bulk and \textit{nutational} spin-pumping contributions to the nutational damping parameter $\alpha_\eta = \alpha_{\eta}^{\text{b}} + \alpha_\eta^\text{nsp}$. A key point is that the $\alpha_\eta^\text{nsp}$-term qualitatively changes the spin pumping response: it is precisely this contribution that generates a nonzero pumped dc spin current that scales as the square of the quadratic precession frequency, our second main result, which we now describe in detail.

\textit{Nutational spin pumping}---
In the LLG equation, the contribution of the dissipative pumped spin current $\vec J_s^\text{dis}$ enters as:
\begin{align} \nonumber
    \dot{\vec m} =& -\gamma \mu_0 \vec m \times \vec H_\text{eff} + \alpha^\text{b} \vec m \times \dot{\vec m} + \eta \vec m \times   \ddot{\vec m}
    \\ \label{eq:llg_pumped}
    &+ \eta \alpha_\eta^\text{b} \vec m \times ( \vec m \times  \ddot{\vec m}) + \dotvec  
    m_\text{pumped}^{\text{diss}},
\end{align}
where the rate of change in magnetization due to the loss of the dissipative spin current ($\vec J_s^{\text{dis}} $) to the adjacent conductors is $\dotvec m_\text{pumped}^{\text{diss}} = \gamma  \vec J_s^{\text{dis}} / \mathcal V M_s$, where $\mathcal V$ is the volume of the FM, and $M_s$ is the saturation magnetization.
In Eq.~\eqref{eq:llg_pumped}, $\alpha^\text{b}$ and $\alpha_\eta^\text{b}$ represent the bulk contributions to Gilbert damping and nutational damping, respectively. 

The spin current $\vec J_s^{\text{dis}}$ scales with the interface area $A$, allowing us to define the average spin current density as $\vec j_s^{\text{dis}} = \vec J_s^{\text{dis}} / A$. For a ferromagnetic slab of thickness $d_F$ and volume $\mathcal V = A d_F$, the spin-pumping-induced change of magnetization satisfies $M_s  \dotvec m_\text{pumped}^{\text{diss}}  = \gamma \vec J_s^{\text{dis}} / \mathcal V =\gamma \vec j_s^{\text{dis}} / d_F$; hence, it scales as $1 / d_F$. In other words, since the total magnetic moment $\mathcal V M_s \vec m$ scales with the total volume of the FM, while the interfacial spin current scales with $A$, we expect the nutational damping to be large in ferromagnetic thin films (small $d_F$), similar to spin-pumping-enhanced Gilbert damping. 

The time-averaged pumped current in an adjacent NM can be expressed generally through the dissipative terms in the LLG equation: 
\begin{align}
\label{eq:Isp}
    \vec j_s^{\text{dis}}   =
    \frac{\hbar g_r^{\uparrow \downarrow}}{4 \pi } \vec m \times \dotvec  m 
    + 
    \frac{\hbar g_\eta}{4 \pi } \eta \, \vec m \times (\vec m \times \ddot{\vec m}),
\end{align}
where $g_r^{\uparrow \downarrow}$ is the real part of the mixing conductance, and $g_\eta$ is the conductance that we introduce to characterize the transfer of a nutationally pumped spin current into the NM. The reactive term $\sim \vec m \times \ddot{\vec{m}}$ produces an oscillating spin current that vanishes when the spin current is averaged over time in the adjacent NM.

Inserting the spin current of Eq.~\eqref{eq:Isp} into the LLG equation in Eq.~\eqref{eq:llg_pumped}, we obtain expressions for the interface damping parameters in terms of conductances:
\begin{align}
    \alpha^\text{sp} = \frac{\hbar \gamma g_r^{\uparrow \downarrow}}{4 \pi M_s d_F} \, , &&
     \alpha^\text{nsp}_\eta = \frac{\hbar \gamma g_\eta}{4 \pi M_s d_F} \, .
\end{align}
Nutational spin pumping and the parameter $g_\eta$ can be measured in several ways. An indirect yet precise measure is via the resonance linewidth. As established \cite{tserkovnyakNonlocalMagnetizationDynamics2005}, the thickness dependence of the Gilbert damping parameter $\alpha$ that describes the width of the FMR resonance line determines the conventional spin-mixing conductance $g_r^{\uparrow \downarrow}$. Similarly, the thickness dependence of the nutational resonance linewidth determines the sum $g_r^{\uparrow \downarrow}+g_\eta$. These two measurements, the FMR linewidth and the nutational linewidth, will then determine the nutational spin pumping conductance $g_\eta$. 
A direct measurement protocol for the conventional and nutational output spin current is described in the following paragraphs.

To find the pumped spin current to the lowest order in the deviation of the magnetization from its equilibrium value, we assume a harmonic driving field $h_x(t) = A_x \cos ( \Omega t)$ and $h_y(t) = A_y \cos ( \Omega t + \phi_h) $, where $A_x, A_y>0$ are the amplitudes, $\Omega > 0$ is the driving frequency and $\phi_h$ is a phase controlling the polarization of the field.
It will be convenient to define the circular components $A_\pm = A_x \pm \iu \e^{\iu \phi_y} A_y$.
We expand the magnetization vector $\vec m$  around the equilibrium direction
$\vec z$ as $ \vec m \simeq  \vec z + \epsilon \delta \vec m^{(1)}_i + \epsilon^2 \delta \vec m^{(2)}_i$, where $\epsilon$ is a small parameter.
\footnote{In the SM~\cite{SeeSupplementalMaterial}, we also consider the situation where the static field is anti-aligned to the laboratory $z$ axis.}
We focus on the dc spin current polarized along the $z$-axis, $j_{s, z}^{\text{dis}} = j_{s, z}^{\text{sp}} + j_{s, z}^{\text{nsp}}$, where $j_{s, z}^{\text{sp}}$ denotes standard spin pumping linear in $\omega$ and $j_{s, z}^{\text{nsp}}$ is the nutational spin pumping quadratic in $\omega$.

The conventional dc spin pumping term is linear in $\Omega$ and quadratic in $\epsilon$.
In contrast, the contribution from the dissipative nutation produces a dc term {\it quadratic} in $\omega$ and quadratic in $\epsilon$.
We set $\epsilon = 1$ in the results below.
We express the spin current in terms of the circular basis and insert the susceptibilities via the relation $\delta m_\pm = \chi_\pm h_\pm$:
\begin{subequations}
\label{eq:phen_j}
\begin{align} 
     j_{s, z}^{\text{sp}} &=  
     c  g_r^{\uparrow \downarrow} \Omega
    (|\chi_+(\Omega)|^2 |A_+|^2 - |\chi_-(\Omega)|^2 |A_-|^2 ),
    \\
    \label{eq:phen_jz2}
        j_{s, z}^{\text{nsp}} &= - 
        c \eta g_\eta  \Omega^2
        (|\chi_+(\Omega)|^2 |A_+|^2 + |\chi_-(\Omega)|^2 |A_-|^2 ),
\end{align}
\end{subequations}
where $c = \hbar / 16 \pi$ is a constant.
The general derivation performed in the SM also reveals ac terms in the nutational spin current with a frequency dependence of $2 \Omega$; however, these terms are typically small and are ignored here.
We emphasize that although spin pumping was analyzed at the nutational scale in Ref.~\onlinecite{mondalSpinPumpingTerahertz2021}, the underlying physics remains conventional spin pumping, with the pumped response scaling linearly with the driving frequency.

The nutational contribution $j^\text{nsp}_{s, z}$ should be experimentally separable from the conventional term $j^\text{sp}_{s, z}$ by its distinctive quadratic frequency scaling. 
A particularly clean extraction follows from assuming a right-handed circular field $\phi_h = - \pi/2$ and $A_x = A_y = A$, such that $A_- = 0$ and $A_+ = 2A$.
We extract the integrated spin current by integrating over a frequency range close to the precessional frequency $\omega_H$, and take the slope of this with respect to $\omega_H$~\cite{SeeSupplementalMaterial}:
\begin{align}
\label{eq:meas}
    \frac{\d}{\d \omega_H} \int \d\Omega \, j_{s, z}  
    \simeq -\frac{\hbar g_\eta \eta (\gamma \mu_0)^2 A^2}{4 \alpha }.
\end{align}
In Eq.~\eqref{eq:meas}, we approximated the Lorentzian distributions arising from the susceptibilities as delta functions close to the precession frequency $\omega_H$, valid when $\alpha \ll 1$.
The signal in Eq.\eqref{eq:meas} is present only for nonzero $g_\eta$, and establishes a clear way to measure nutational spin pumping without interference from conventional spin pumping.
We emphasize that what we provide here is a way to measure the \textit{nutational} spin pumping $j_{s, z}^\text{nsp}$ at the FMR frequency. 
A second direct measurement of nutational spin pumping involves analyzing the total output current near the FMR and NR frequencies; see the SM.

\textit{Microscopic scattering formulation}---
Our third main result is a microscopic theory with explicit expressions for nutational spin pumping. To obtain microscopic expressions for the nutational spin pumping-induced damping parameter $\alpha_\eta^\text{nsp}$, we compute the pumped spin current from an insulating FM ($x> 0$) in contact with a metallic NM ($x<0$) connected to a lead, see Fig.~\ref{fig:1} for a sketch of the system. 
This approach straightforwardly generalizes to yield the corresponding expressions for nutational spin pumping in other systems, including metallic NM–FM–NM systems.

The time-dependent Hamiltonian of the system is 
\begin{align}
    H = H_0(x, \vrho) + \Theta(x) [V_0 +\Delta \vec m \cdot \bm{\sigma}],
\end{align}
where $H_0(x, \boldsymbol \rho) = -\frac{\hbar^2}{2m} \partial_x^2 + H_\perp(\boldsymbol \rho)$ represents the nonmagnetic Hamiltonian for a free particle in the $x$-direction, with the transverse coordinate $\boldsymbol \rho$ that results in transverse waveguide modes $n$. $\Theta(x)$ is the Heaviside function. 
$\Delta$ is the exchange splitting and $V_0$ is the charge potential barrier, both active only inside the FM.
For clarity in this work, we assume that $V_0$ and $\Delta$ are homogeneous within the FM. 
The magnetization direction of FM is time-dependent and is given by $\vec m = (\sin \theta \cos \phi, \sin \theta \sin \phi, \cos \theta)$, where $\theta$ and $\phi$ are the polar and azimuthal angles. 
We assume simple precession, i.e., $\phi = \omega t$ and $\theta$ is a constant, and perform a unitary transformation $U = \e^{- \iu \omega t \sigma_z / 2}$ into a rotating frame, resulting in a transformed Hamiltonian, 
$\tilde H =  U^\dag  H  U - \iu U^\dag \partial_t U.$
In the case of simple precession, the rotated Hamiltonian becomes time-independent:    
$
    \tilde{H} = \tilde H_0 + \Theta(x) [V_0 +\Delta \mo \cdot \bm{\sigma}],
$
where $\tilde H_0 = H_0 - \frac{\hbar \omega}{2} \sigma_z$ is the normal metal Hamiltonian in the rotating frame.
In the rotating frame, the exchange splitting in the FM is along a constant unit vector $\mo = (\sin \theta, 0, \cos \theta)$.
 Unitary rotation $U$ renders the problem time independent; we then solve the Schrödinger equation in each region. 

The spin current in the normal metal follows from integrating the spin‑current density over the transverse coordinate,
\begin{align} \label{eq:jz_gen}
    J_{s, z} =  \frac{\hbar^2}{2 m } \int \d\vrho \, \Im{\langle { \Psi}^\dag(x, \vrho, t) \sigma_z \partial_x {  \Psi}(x, \vrho, t) \rangle },
\end{align}
where $\Psi(x)$ is the field operator in the lab frame, which is related to the rotating frame field operator as $\Psi = U {\tilde \Psi}$, and $\langle \ldots \rangle$ denotes the expectation value of the operator.
In the lab frame, we express the spin current through the rotating‑frame reflection matrix $\tilde r$ (see the SM~\cite{SeeSupplementalMaterial} for the full derivation).
Crucially, the spin current depends on the driving frequency $\omega$ in two ways: through the energy argument of the Fermi-Dirac function $f$ \footnote{Due to the unitary transformation into a rotating frame, there is an effective spin accumulation in the rotating frame.} and in the reflection matrix $\tilde r$. Unlike conventional spin pumping, the latter dependence implies that we have to go beyond the frozen scattering matrix approximation to compute nutational spin pumping; e.g., conventional adiabatic pumping, as in Ref.~\onlinecite{brouwerScatteringApproachParametric1998}, does not capture nutational spin pumping. 

Expanding the spin current to first order in the FMR frequency $\omega$, we reproduce the conventional result of linear spin pumping in terms of the real part of the mixing conductance $J_{s, z}^{\text{sp}} = - (4\pi)^{-1} \hbar \omega g^r_{\uparrow \downarrow} \sin^2  \theta$, where the negative sign is because the spins flow to the left into the normal metal in Fig.\ \ref{fig:1}.
The second-order nutational spin-pumping contribution depends on contributions beyond the frozen scattering matrix and becomes
\begin{align} 
    J_{s, z}^{\text{nsp}} = \frac{\hbar \omega^2}{4 \pi} \sum_{s \sigma nm} s \sigma \Re{(\tilde r_{\sigma s}^{mn})^* \partial_\omega \tilde r_{\sigma s}^{mn}}_F, 
\end{align}
where $F$ denotes evaluation at the Fermi energy and at $\omega = 0$, $s, \sigma$ are spin indices, and $n,m$ are transverse waveguide mode indices.
 
\begin{figure}[t]
    \centering
    \begin{subfigure}{0.492\linewidth}
        \includegraphics[width=\linewidth]{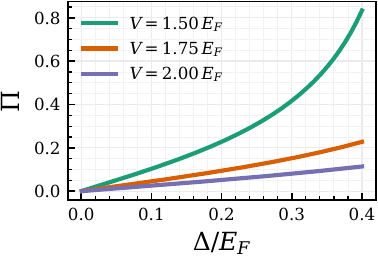}
        \caption{}
    \end{subfigure}
    \hfill
    \begin{subfigure}{0.492\linewidth}
        \includegraphics[width=\linewidth]{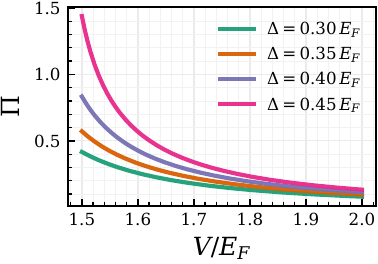}
        \caption{}
    \end{subfigure}
    \caption{The strength of the nutational damping $\Pi$ as a function of (a) $\Delta$ and (b) $V$.}
    \label{fig:2}
\end{figure}

For illustration, we consider a single transverse mode and obtain an expression for the second-order spin current as
\begin{align} \label{eq:ratio_scatt}
    J^{\text{nsp}}_{s, z} / J^{\text{sp}}_{s, z} = \frac{\hbar \omega}{E_F}  \Pi(V , \Delta, E_F) \cos{\theta},
\end{align}
where $\Pi$ denotes a complicated function that depends on the system parameters; its complete expression, which offers little additional insight, is given in the SM.
$\Pi$ vanishes for $\Delta = 0$ and grows linearly with $\Delta$ in the weak-exchange limit.
Fig.~\ref{fig:2} plots $\Pi$ for several potentials and exchange splittings.
In the long-penetration regime inside the FM, $V  \sim \Delta + E_F$, we find $\Pi \sim 1$, implying that the second-order spin current is suppressed only by $\hbar \omega / E_F$ relative to the first-order term. Consequently, the quadratic term rapidly gains weight as $\omega$ increases and can become significant in the high-frequency regime. This trend for the {\it dissipative} nutational spin-pumping echoes earlier theoretical results for the {\it reactive} nutational parameter $\eta$
in metallic ferromagnets~\cite{kikuchiSpinDynamicsInertia2015}. 
To obtain an expression for the dissipative nutational parameter $\alpha_\eta^\text{nsp}$, consider the ratio of the first-order and second-order spin currents, derived using a phenomenological approach and valid for a harmonic drive, in Eq.~\eqref{eq:phen_j}:
$ \label{eq:ratio_phen}
    J^{\text{nsp}}_{s, z} / J^{\text{sp}}_{s, z} =
    - \alpha_\eta^\text{nsp} \eta \omega / \alpha^\text{sp}
$.
Comparing this result with the result obtained from the scattering system, Eq.~\eqref{eq:ratio_scatt}, yields an expression for the nutational spin-pumping parameter:
\begin{align}
    \frac{\alpha_\eta^\text{nsp} }{ \alpha^\text{sp}} = 
    - \frac{\hbar}{\eta E_F } \Pi(V, \Delta, E_F).
\end{align}
Here, we set the $\cos \theta$ factor to unity since we consider linear response, i.e., small angles $\theta$.

Finally, we estimate the size of the parameter $\alpha_\eta$.
Considering the regime where $\Pi \sim 1$, and assuming that the nutational precession time is $\eta \sim \hbar / E_F$, as found in a simple model of a conducting FM~\cite{kikuchiSpinDynamicsInertia2015}, we estimate $\alpha_\eta^\text{nsp} / \alpha^\text{sp} $ to be of order one. 
This estimate implies that nutational spin pumping in thin films is essential for the linewidth of the nutational excitation expressed in Eq.~\eqref{eq:resonances}.
We expect the nutational spin pumping effect to become important as the frequency approaches the ultrahigh-frequency $\omega_\eta$, analogous to the emergence of standard inertial effects at these frequencies.
Since $\eta \sim \hbar / E_F$ \cite{kikuchiSpinDynamicsInertia2015}, the nutational effects are relatively stronger for systems with lower Fermi energy, such as semiconductors. 

A complete treatment of realistic band structures for computing $\eta$ and $g_\eta$ lies beyond the scope of this work. Nevertheless, we anticipate that both parameters can be substantially enhanced in systems where spin-opposite bands intersect, as such crossings promote dynamic spin-flip scattering.
This expectation aligns with experimental findings, where $\eta$ in typical ferromagnets lies in the picosecond range \cite{neerajInertialSpinDynamics2021, unikandanunniInertialSpinDynamics2022, deMagneticNutationTransient2025}, far exceeding some theoretical estimates for simple-band models predicting femtosecond to sub-picosecond values \cite{kikuchiSpinDynamicsInertia2015, thonigMagneticMomentInertia2017}. Thus, nutational spin pumping might become relevant already in the picosecond (i.e., \si{\tera \hertz}) range.

\textit{Concluding remarks}---
In summary, we identify a dissipative nutational torque in the inertial LLG equation that governs the lifetime of the nutational mode and becomes relevant at terahertz frequencies. 
Nutational spin pumping is accessable via FMR and nutational-resonance experiments, \si{\tera\hertz} emission spectroscopy, spin-rectification measurements, and inverse spin Hall effect measurements in ferromagnet–nonmagnetic-metal heterostructures.
Beyond shaping magnetization dynamics, the same term drives quadratic spin pumping under ac excitation, generating a dc spin current with $z$-polarized spin. At and around FMR frequencies, total nutational spin pumping is proportional to the external dc magnetic field, whereas conventional spin pumping is independent of the same field. 
Accordingly, isolating the nutational spin signal amounts to evaluating the slope with respect to the static field.
The spin currents are accessible experimentally via inverse spin Hall effect measurements in ferromagnet–heavy-metal heterostructures.
Our results move spintronics beyond the Gilbert paradigm and open the route to ultrafast spin-current generation for next-generation devices.

\begin{acknowledgments}
{\it Acknowledgments}---
We thank Chiara Ciccarelli and Ajinkya Punjal for useful comments.
The Research Council of Norway supported this work through its Centers of Excellence funding scheme, Project No. 262633, ''QuSpin''.
\end{acknowledgments}

\bibliography{main}

@article{bastardisMagnetizationNutationInduced2018a,
  title = {Magnetization Nutation Induced by Surface Effects in Nanomagnets},
  author = {Bastardis, R. and Vernay, F. and Kachkachi, H.},
  year = 2018,
  month = oct,
  journal = {Phys. Rev. B},
  volume = {98},
  number = {16},
  pages = {165444},
  issn = {2469-9950, 2469-9969},
  doi = {10.1103/PhysRevB.98.165444},
  urldate = {2026-02-17},
  langid = {english}
}

@article{bhattacharjeeAtomisticSpinDynamic2012a,
  title = {Atomistic {{Spin Dynamic Method}} with Both {{Damping}} and {{Moment}} of {{Inertia Effects Included}} from {{First Principles}}},
  author = {Bhattacharjee, Satadeep and Nordstr{\"o}m, Lars and Fransson, Jonas},
  year = 2012,
  month = jan,
  journal = {Phys. Rev. Lett.},
  volume = {108},
  number = {5},
  pages = {057204},
  issn = {0031-9007, 1079-7114},
  doi = {10.1103/PhysRevLett.108.057204},
  urldate = {2026-02-17},
  copyright = {http://link.aps.org/licenses/aps-default-license},
  langid = {english}
}

@article{brouwerScatteringApproachParametric1998,
  title = {Scattering Approach to Parametric Pumping},
  author = {Brouwer, P. W.},
  year = 1998,
  month = oct,
  journal = {Phys. Rev. B},
  volume = {58},
  number = {16},
  pages = {R10135-R10138},
  issn = {0163-1829, 1095-3795},
  doi = {10.1103/PhysRevB.58.R10135},
  urldate = {2026-01-16},
  copyright = {http://link.aps.org/licenses/aps-default-license},
  langid = {english}
}

@article{cherkasskiiDispersionRelationNutation2021,
  title = {Dispersion Relation of Nutation Surface Spin Waves in Ferromagnets},
  author = {Cherkasskii, Mikhail and Farle, Michael and Semisalova, Anna},
  year = 2021,
  month = may,
  journal = {Phys. Rev. B},
  volume = {103},
  number = {17},
  pages = {174435},
  issn = {2469-9950, 2469-9969},
  doi = {10.1103/PhysRevB.103.174435},
  urldate = {2026-01-19},
  langid = {english}
}

@article{cherkasskiiNutationResonanceFerromagnets2020,
  title = {Nutation Resonance in Ferromagnets},
  author = {Cherkasskii, Mikhail and Farle, Michael and Semisalova, Anna},
  year = 2020,
  month = nov,
  journal = {Phys. Rev. B},
  volume = {102},
  number = {18},
  pages = {184432},
  issn = {2469-9950, 2469-9969},
  doi = {10.1103/PhysRevB.102.184432},
  urldate = {2026-01-08},
  langid = {english}
}

@article{cherkasskiiTheoryInertialSpin2022,
  title = {Theory of Inertial Spin Dynamics in Anisotropic Ferromagnets},
  author = {Cherkasskii, Mikhail and Barsukov, Igor and Mondal, Ritwik and Farle, Michael and Semisalova, Anna},
  year = 2022,
  month = aug,
  journal = {Phys. Rev. B},
  volume = {106},
  number = {5},
  pages = {054428},
  issn = {2469-9950, 2469-9969},
  doi = {10.1103/PhysRevB.106.054428},
  urldate = {2026-01-08},
  langid = {english}
}

@article{ciorneiMagnetizationDynamicsInertial2011,
  title = {Magnetization Dynamics in the Inertial Regime: {{Nutation}} Predicted at Short Time Scales},
  shorttitle = {Magnetization Dynamics in the Inertial Regime},
  author = {Ciornei, M.-C. and Rub{\'i}, J. M. and Wegrowe, J.-E.},
  year = 2011,
  month = jan,
  journal = {Phys. Rev. B},
  volume = {83},
  number = {2},
  pages = {020410},
  issn = {1098-0121, 1550-235X},
  doi = {10.1103/PhysRevB.83.020410},
  urldate = {2025-06-02},
  copyright = {http://link.aps.org/licenses/aps-default-license},
  langid = {english}
}

@article{deMagneticNutationTransient2025,
  title = {Magnetic Nutation: {{Transient}} Separation of Magnetization from Its Angular Momentum},
  shorttitle = {Magnetic Nutation},
  author = {De, Anulekha and Schlegel, Julius and Lentfert, Akira and Scheuer, Laura and Stadtm{\"u}ller, Benjamin and Pirro, Philipp and Von Freymann, Georg and Nowak, Ulrich and Aeschlimann, Martin},
  year = 2025,
  month = jan,
  journal = {Phys. Rev. B},
  volume = {111},
  number = {1},
  pages = {014432},
  issn = {2469-9950, 2469-9969},
  doi = {10.1103/PhysRevB.111.014432},
  urldate = {2025-04-25},
  langid = {english}
}

@article{fahnleGeneralizedGilbertEquation2011,
  title = {Generalized {{Gilbert}} Equation Including Inertial Damping: {{Derivation}} from an Extended Breathing {{Fermi}} Surface Model},
  shorttitle = {Generalized {{Gilbert}} Equation Including Inertial Damping},
  author = {F{\"a}hnle, Manfred and Steiauf, Daniel and Illg, Christian},
  year = 2011,
  month = nov,
  journal = {Phys. Rev. B},
  volume = {84},
  number = {17},
  pages = {172403},
  issn = {1098-0121, 1550-235X},
  doi = {10.1103/PhysRevB.84.172403},
  urldate = {2025-12-13},
  copyright = {http://link.aps.org/licenses/aps-default-license},
  langid = {english}
}

@article{ghoshTheoryTensorialMagnetic2024,
  title = {Theory of Tensorial Magnetic Inertia in Terahertz Spin Dynamics},
  author = {Ghosh, Subhadip and Cherkasskii, Mikhail and Barsukov, Igor and Mondal, Ritwik},
  year = 2024,
  month = nov,
  journal = {Phys. Rev. B},
  volume = {110},
  number = {17},
  pages = {174430},
  issn = {2469-9950, 2469-9969},
  doi = {10.1103/PhysRevB.110.174430},
  urldate = {2026-02-17},
  langid = {english}
}

@article{gilbertClassicsMagneticsPhenomenological2004,
  title = {Classics in {{Magnetics A Phenomenological Theory}} of {{Damping}} in {{Ferromagnetic Materials}}},
  author = {Gilbert, T.L.},
  year = 2004,
  month = nov,
  journal = {IEEE Trans. Magn.},
  volume = {40},
  number = {6},
  pages = {3443--3449},
  issn = {0018-9464},
  doi = {10.1109/TMAG.2004.836740},
  urldate = {2026-02-17},
  copyright = {https://ieeexplore.ieee.org/Xplorehelp/downloads/license-information/IEEE.html},
  langid = {english}
}

@article{johnsenDynamicallyGeneratedSpin2025,
  title = {Dynamically Generated Spin Interactions and Nutational Spin Inertia in Normal Metal--Ferromagnet Heterostructures},
  author = {Johnsen, Christian Svingen and Sudb{\o}, Asle},
  year = 2025,
  month = apr,
  journal = {Phys. Rev. B},
  volume = {111},
  number = {14},
  pages = {144423},
  issn = {2469-9950, 2469-9969},
  doi = {10.1103/PhysRevB.111.144423},
  urldate = {2026-02-17},
  langid = {english}
}

@article{khodadadiEnhancedSpinPumping2017,
  title = {Enhanced Spin Pumping near a Magnetic Ordering Transition},
  author = {Khodadadi, Behrouz and Mohammadi, Jamileh Beik and Mewes, Claudia and Mewes, Tim and Manno, M. and Leighton, C. and Miller, Casey W.},
  year = 2017,
  month = aug,
  journal = {Phys. Rev. B},
  volume = {96},
  number = {5},
  pages = {054436},
  issn = {2469-9950, 2469-9969},
  doi = {10.1103/PhysRevB.96.054436},
  urldate = {2025-06-10},
  copyright = {https://link.aps.org/licenses/aps-default-license},
  langid = {english}
}

@article{kikuchiSpinDynamicsInertia2015,
  title = {Spin Dynamics with Inertia in Metallic Ferromagnets},
  author = {Kikuchi, Toru and Tatara, Gen},
  year = 2015,
  month = nov,
  journal = {Phys. Rev. B},
  volume = {92},
  number = {18},
  pages = {184410},
  issn = {1098-0121, 1550-235X},
  doi = {10.1103/PhysRevB.92.184410},
  urldate = {2025-12-08},
  copyright = {http://link.aps.org/licenses/aps-default-license},
  langid = {english}
}

@article{kimelInertiadrivenSpinSwitching2009,
  title = {Inertia-Driven Spin Switching in Antiferromagnets},
  author = {Kimel, A. V. and Ivanov, B. A. and Pisarev, R. V. and Usachev, P. A. and Kirilyuk, A. and Rasing, {\relax Th}.},
  year = 2009,
  month = oct,
  journal = {Nat. Phys.},
  volume = {5},
  number = {10},
  pages = {727--731},
  issn = {1745-2473, 1745-2481},
  doi = {10.1038/nphys1369},
  urldate = {2025-06-02},
  langid = {english}
}

@article{landauTheoryDispersionMagnetic1935,
  title = {On the Theory of the Dispersion of Magnetic Permeability in Ferromagnetic Bodies},
  author = {Landau, L and Lifshitz, E},
  year = 1935,
  journal = {Phys. Z. Sowjetunion},
  volume = {8},
  pages = {153--169},
  langid = {english}
}

@article{liInertialTermsMagnetization2015,
  title = {Inertial Terms to Magnetization Dynamics in Ferromagnetic Thin Films},
  author = {Li, Y. and Barra, A.-L. and Auffret, S. and Ebels, U. and Bailey, W. E.},
  year = 2015,
  month = oct,
  journal = {Phys. Rev. B},
  volume = {92},
  number = {14},
  pages = {140413},
  issn = {1098-0121, 1550-235X},
  doi = {10.1103/PhysRevB.92.140413},
  urldate = {2026-01-16},
  copyright = {http://link.aps.org/licenses/aps-default-license},
  langid = {english}
}

@article{lomonosovAnatomyInertialMagnons2021,
  title = {Anatomy of Inertial Magnons in Ferromagnetic Nanostructures},
  author = {Lomonosov, Alexey M. and Temnov, Vasily V. and Wegrowe, Jean-Eric},
  year = 2021,
  month = aug,
  journal = {Phys. Rev. B},
  volume = {104},
  number = {5},
  pages = {054425},
  issn = {2469-9950, 2469-9969},
  doi = {10.1103/PhysRevB.104.054425},
  urldate = {2026-01-19},
  langid = {english}
}

@article{mondalGeneralisationGilbertDamping2018,
  title = {Generalisation of {{Gilbert}} Damping and Magnetic Inertia Parameter as a Series of Higher-Order Relativistic Terms},
  author = {Mondal, Ritwik and Berritta, Marco and Oppeneer, Peter M},
  year = 2018,
  month = jul,
  journal = {J. Phys.: Condens. Matter},
  volume = {30},
  number = {26},
  pages = {265801},
  issn = {0953-8984, 1361-648X},
  doi = {10.1088/1361-648X/aac5a2},
  urldate = {2026-01-16},
  langid = {english}
}

@article{mondalInertialEffectsUltrafast2023,
  title = {Inertial Effects in Ultrafast Spin Dynamics},
  author = {Mondal, Ritwik and R{\'o}zsa, Levente and Farle, Michael and Oppeneer, Peter M. and Nowak, Ulrich and Cherkasskii, Mikhail},
  year = 2023,
  month = aug,
  journal = {J. Magn. Magn. Mater.},
  volume = {579},
  pages = {170830},
  issn = {03048853},
  doi = {10.1016/j.jmmm.2023.170830},
  urldate = {2025-06-20},
  langid = {english}
}

@article{mondalInertialSpinWaves2022,
  title = {Inertial Spin Waves in Ferromagnets and Antiferromagnets},
  author = {Mondal, Ritwik and R{\'o}zsa, Levente},
  year = 2022,
  month = oct,
  journal = {Phys. Rev. B},
  volume = {106},
  number = {13},
  pages = {134422},
  issn = {2469-9950, 2469-9969},
  doi = {10.1103/PhysRevB.106.134422},
  urldate = {2026-01-16},
  langid = {english}
}

@article{mondalSpinPumpingTerahertz2021,
  title = {Spin Pumping at Terahertz Nutation Resonances},
  author = {Mondal, Ritwik and Kamra, Akashdeep},
  year = 2021,
  month = dec,
  journal = {Phys. Rev. B},
  volume = {104},
  number = {21},
  pages = {214426},
  issn = {2469-9950, 2469-9969},
  doi = {10.1103/PhysRevB.104.214426},
  urldate = {2025-06-20},
  langid = {english}
}

@article{neerajInertialSpinDynamics2021,
  title = {Inertial Spin Dynamics in Ferromagnets},
  author = {Neeraj, Kumar and Awari, Nilesh and Kovalev, Sergey and Polley, Debanjan and Zhou Hagstr{\"o}m, Nanna and Arekapudi, Sri Sai Phani Kanth and Semisalova, Anna and Lenz, Kilian and Green, Bertram and Deinert, Jan-Christoph and Ilyakov, Igor and Chen, Min and Bawatna, Mohammed and Scalera, Valentino and {d'Aquino}, Massimiliano and Serpico, Claudio and Hellwig, Olav and Wegrowe, Jean-Eric and Gensch, Michael and Bonetti, Stefano},
  year = 2021,
  month = feb,
  journal = {Nat. Phys.},
  volume = {17},
  number = {2},
  pages = {245--250},
  issn = {1745-2473, 1745-2481},
  doi = {10.1038/s41567-020-01040-y},
  urldate = {2025-06-03},
  langid = {english}
}

@article{oliveDeviationLandauLifshitzGilbertEquation2015,
  title = {Deviation from the {{Landau-Lifshitz-Gilbert}} Equation in the Inertial Regime of the Magnetization},
  author = {Olive, E. and Lansac, Y. and Meyer, M. and Hayoun, M. and Wegrowe, J.-E.},
  year = 2015,
  month = jun,
  journal = {J. Appl. Phys.},
  volume = {117},
  number = {21},
  pages = {213904},
  issn = {0021-8979, 1089-7550},
  doi = {10.1063/1.4921908},
  urldate = {2026-01-08},
  langid = {english}
}

@article{oliveFerromagneticResonanceInertial2012,
  title = {Beyond Ferromagnetic Resonance: {{The}} Inertial Regime of the Magnetization},
  shorttitle = {Beyond Ferromagnetic Resonance},
  author = {Olive, E. and Lansac, Y. and Wegrowe, J.-E.},
  year = 2012,
  month = may,
  journal = {Appl. Phys. Lett.},
  volume = {100},
  number = {19},
  pages = {192407},
  issn = {0003-6951, 1077-3118},
  doi = {10.1063/1.4712056},
  urldate = {2026-01-08},
  langid = {english}
}

@article{quarentaBathInducedSpinInertia2024,
  title = {Bath-{{Induced Spin Inertia}}},
  author = {Quarenta, Mario Gaspar and Tharmalingam, Mithuss and Ludwig, Tim and Yuan, H. Y. and Karwacki, Lukasz and Verstraten, Robin C. and Duine, Rembert A.},
  year = 2024,
  month = sep,
  journal = {Phys. Rev. Lett.},
  volume = {133},
  number = {13},
  pages = {136701},
  issn = {0031-9007, 1079-7114},
  doi = {10.1103/PhysRevLett.133.136701},
  urldate = {2024-12-22},
  langid = {english}
}

@article{rodriguezSpinInertiaAutoOscillations2024,
  title = {Spin {{Inertia}} and {{Auto-Oscillations}} in {{Ferromagnets}}},
  author = {Rodriguez, Rodolfo and Cherkasskii, Mikhail and Jiang, Rundong and Mondal, Ritwik and Etesamirad, Arezoo and Tossounian, Allison and Ivanov, Boris A. and Barsukov, Igor},
  year = 2024,
  month = jun,
  journal = {Phys. Rev. Lett.},
  volume = {132},
  number = {24},
  pages = {246701},
  issn = {0031-9007, 1079-7114},
  doi = {10.1103/PhysRevLett.132.246701},
  urldate = {2026-01-16},
  langid = {english}
}

@misc{SeeSupplementalMaterial,
  title = {See Supplemental Material at [URL will be inserted by publisher] for additional details on the derivation and linearization of the generalized inertial LLG equation, the pumped spin current, and scattering theory},
  year = {2026}
}

@article{thonigMagneticMomentInertia2017,
  title = {Magnetic Moment of Inertia within the Torque-Torque Correlation Model},
  author = {Thonig, Danny and Eriksson, Olle and Pereiro, Manuel},
  year = 2017,
  month = apr,
  journal = {Sci. Rep.},
  volume = {7},
  number = {1},
  pages = {931},
  issn = {2045-2322},
  doi = {10.1038/s41598-017-01081-z},
  urldate = {2026-01-08},
  langid = {english}
}

@article{tiberkevichNonlinearPhenomenologicalModel2007,
  title = {Nonlinear Phenomenological Model of Magnetic Dissipation for Large Precession Angles: {{Generalization}} of the {{Gilbert}} Model},
  shorttitle = {Nonlinear Phenomenological Model of Magnetic Dissipation for Large Precession Angles},
  author = {Tiberkevich, Vasil and Slavin, Andrei},
  year = 2007,
  month = jan,
  journal = {Phys. Rev. B},
  volume = {75},
  number = {1},
  pages = {014440},
  issn = {1098-0121, 1550-235X},
  doi = {10.1103/PhysRevB.75.014440},
  urldate = {2026-04-15},
  copyright = {http://link.aps.org/licenses/aps-default-license},
  langid = {english}
}

@article{titovNutationSpinWaves2022,
  title = {Nutation Spin Waves in Ferromagnets},
  author = {Titov, Sergei V. and Dowling, William J. and Kalmykov, Yuri P. and Cherkasskii, Mikhail},
  year = 2022,
  month = jun,
  journal = {Phys. Rev. B},
  volume = {105},
  number = {21},
  pages = {214414},
  issn = {2469-9950, 2469-9969},
  doi = {10.1103/PhysRevB.105.214414},
  urldate = {2026-01-19},
  langid = {english}
}

@article{tserkovnyakEnhancedGilbertDamping2002,
  title = {Enhanced {{Gilbert Damping}} in {{Thin Ferromagnetic Films}}},
  author = {Tserkovnyak, Yaroslav and Brataas, Arne and Bauer, Gerrit E. W.},
  year = 2002,
  month = feb,
  journal = {Phys. Rev. Lett.},
  volume = {88},
  number = {11},
  pages = {117601},
  issn = {0031-9007, 1079-7114},
  doi = {10.1103/PhysRevLett.88.117601},
  urldate = {2025-06-10},
  copyright = {http://link.aps.org/licenses/aps-default-license},
  langid = {english}
}

@article{tserkovnyakNonlocalMagnetizationDynamics2005,
  title = {Nonlocal Magnetization Dynamics in Ferromagnetic Heterostructures},
  author = {Tserkovnyak, Yaroslav and Brataas, Arne and Bauer, Gerrit E. W. and Halperin, Bertrand I.},
  year = 2005,
  month = dec,
  journal = {Rev. Mod. Phys.},
  volume = {77},
  number = {4},
  pages = {1375--1421},
  issn = {0034-6861, 1539-0756},
  doi = {10.1103/RevModPhys.77.1375},
  urldate = {2025-06-10},
  copyright = {http://link.aps.org/licenses/aps-default-license},
  langid = {english}
}

@article{tserkovnyakSpinPumpingMagnetization2002,
  title = {Spin Pumping and Magnetization Dynamics in Metallic Multilayers},
  author = {Tserkovnyak, Yaroslav and Brataas, Arne and Bauer, Gerrit E. W.},
  year = 2002,
  month = dec,
  journal = {Phys. Rev. B},
  volume = {66},
  number = {22},
  pages = {224403},
  issn = {0163-1829, 1095-3795},
  doi = {10.1103/PhysRevB.66.224403},
  urldate = {2025-06-10},
  copyright = {http://link.aps.org/licenses/aps-default-license},
  langid = {english}
}

@article{unikandanunniInertialSpinDynamics2022,
  title = {Inertial {{Spin Dynamics}} in {{Epitaxial Cobalt Films}}},
  author = {Unikandanunni, Vivek and Medapalli, Rajasekhar and Asa, Marco and Albisetti, Edoardo and Petti, Daniela and Bertacco, Riccardo and Fullerton, Eric E. and Bonetti, Stefano},
  year = 2022,
  month = nov,
  journal = {Phys. Rev. Lett.},
  volume = {129},
  number = {23},
  pages = {237201},
  issn = {0031-9007, 1079-7114},
  doi = {10.1103/PhysRevLett.129.237201},
  urldate = {2025-06-03},
  langid = {english}
}

@article{wegroweMagnetizationDynamicsGyromagnetic2012,
  title = {Magnetization Dynamics, Gyromagnetic Relation, and Inertial Effects},
  author = {Wegrowe, J.-E. and Ciornei, M.-C.},
  year = 2012,
  month = jul,
  journal = {Am. J. Phys.},
  volume = {80},
  number = {7},
  pages = {607--611},
  issn = {0002-9505, 1943-2909},
  doi = {10.1119/1.4709188},
  urldate = {2025-06-10},
  langid = {english}
}

@article{winterNutationalSwitchingFerromagnets2022,
  title = {Nutational Switching in Ferromagnets and Antiferromagnets},
  author = {Winter, Lucas and Gro{\ss}enbach, Sebastian and Nowak, Ulrich and R{\'o}zsa, Levente},
  year = 2022,
  month = dec,
  journal = {Phys. Rev. B},
  volume = {106},
  number = {21},
  pages = {214403},
  issn = {2469-9950, 2469-9969},
  doi = {10.1103/PhysRevB.106.214403},
  urldate = {2026-01-19},
  langid = {english}
}
\end{document}


\title{Supplemental Material for \emph{Nutational Spin Pumping and Dissipation}}

\author{Hans Gløckner Giil}
\author{Arne Brataas}
\affiliation{Center for Quantum Spintronics, Department of Physics, Norwegian University of Science and Technology, NO-7491 Trondheim, Norway}
\date{\today}
\maketitle

\tableofcontents

\section{Phenomenological derivation of the inertial LLG equation} \label{sec:sm_phen}
In this section, we present a phenomenological derivation of terms that may appear in the inertial Landau-Lifshitz-Gilbert (LLG) equation, up to second-order temporal variations.
We construct this as a low-energy effective theory, retaining the minimal set of symmetry-allowed terms at each order. The coefficients are independent phenomenological parameters, not terms in a frequency expansion of a single presupposed damping function.
We consider an isotropic system. 
\subsection{Field terms}
All terms in the generalized LLG equation must preserve the magnitude of the magnetization vector $\vec{m}(t)$. Consequently, the most general form of the LLG equation can be expressed as
\begin{align}\label{eq:genLLG_form}
    \dotvec{m} = \vec{m} \times \vec{H}_\text{eff}.
\end{align}
The effective field can contain terms of any order of temporal variations of the magnetization, $\vec H_\text{eff} = \vec H_\text{eff}^{(0)} + \vec H_\text{eff}^{(1)} + \ldots $.
In zeroth order in the rate of change of $\vec{m}$, the effective field $\vec{H}_\text{eff}$ can include terms of the form
\begin{align}
\label{eq:gen_Heff}
    \vec{H}_\text{eff}^{(0)} =  \gamma_1 \vec{H}^{(0)} + \gamma_2 \vec{m} \times \vec{H}^{(0)} + \gamma_3 \vec{m} \times (\vec{m} \times \vec{H}^{(0)}).
\end{align}
Eq. \eqref{eq:gen_Heff} is a general way to write any three-dimensional vector, since $\{\vec{H}^{(0)},  \vec{m} \times \vec{H}^{(0)}, \vec{m} \times (\vec{m} \times \vec{H}^{(0)}\} $ constitutes a basis for three-dimensional space as long as the magnetization and effective fields are non-collinear.
Using the vector triple product identity,
\begin{align} \label{eq:triple_prod}
\vec{A} \times (\vec{B} \times \vec{C}) = \vec{B} (\vec{A} \cdot \vec{C}) - \vec{C} (\vec{A} \cdot \vec{B}),
\end{align}
the $\gamma_3$-term of Eq.~\eqref{eq:gen_Heff} can be rewritten as $\vec{m} (\vec{m} \cdot \vec{H}^{(0)}) - \vec{H}^{(0)}$, where we have used the normalization condition $\vec{m}^2 = 1$.
Substituting this expression back into Eq.~\eqref{eq:genLLG_form}, the first term vanishes due to the cross product with $\vec{m}$, while the second term effectively renormalizes the $\gamma_1$ coefficient of Eq.~\eqref{eq:gen_Heff}, $\gamma_1 \rightarrow \gamma_0$. 
Thus, the most general form of the transverse fields resulting from Eq.~\eqref{eq:gen_Heff} is reduced to the form
\[
\vec{H}_\text{eff} = {\gamma}_0 \vec{H}_0 + {\gamma}_2 \vec{m} \times \vec{H}_0.
\]

Finally, the term proportional to ${\gamma}_2$ can be rewritten in the standard form $ \sim \vec{m} \times \dotvec{m}$ with another renormalization of $\gamma_0$, as is customary when transforming the Landau-Lifshitz (LL) equation into the Landau-Lifshitz-Gilbert (LLG) form.
\subsection{First-order terms}
To first order in the rate of change of $\vec{m}$, the most general form of the effective field is
\begin{align}
    \vec{H}_\text{eff}^{(1)} = \alpha_1 \dotvec{m} + \alpha_2 \vec{m} \times \dotvec{m} + \alpha_3 \vec{m} \times (\vec{m} \times \dotvec{m}).
\end{align}
Using the identity $\vec{m} \times (\vec{m} \times \dotvec{m}) = -\dotvec{m}$, the $\alpha_3$ term simplifies to a renormalization of the $\alpha_1$ term. As a result, the effective field can be written in the reduced form
\[
\vec{H}_\text{eff}^{(1)} = {\alpha}_1 \dotvec{m} + {\alpha}_2 \vec{m} \times \dotvec{m}.
\]
It is also evident that the ${\alpha}_2$ term leads to a renormalization of the gyromagnetic ratio when inserted into Eq.~\eqref{eq:genLLG_form}, again due to the identity $\vec{m} \times (\vec{m} \times \dotvec{m}) = -\dotvec{m}$. Therefore, in the phenomenological derivation presented here, only the dissipative ${\alpha}_1$ term is retained.
\subsection{Second-order terms}
To second order in the first order of rate of change of $\vec{m}$, the allowed terms in the effective field are
\begin{align}
    \vec{H}_\text{eff}^{(2)} = 
    \lambda_1 \underbrace{\dotvec{m} \times \dotvec{m}}_{= 0} 
    + \lambda_2 \dotvec{m} \times (\vec{m} \times \dotvec{m}) 
    + \lambda_3 \underbrace{(\vec{m} \times \dotvec{m}) \times (\vec{m} \times \dotvec{m})}_{= 0} 
    + \ldots
\end{align}
The $\lambda_2$ term can be simplified using the vector identity in Eq. \eqref{eq:triple_prod}:
\[
\dotvec{m} \times (\vec{m} \times \dotvec{m}) = \vec{m} |\dotvec{m}|^2,
\]
which is parallel to $\vec{m}$. When substituted into Eq.~\eqref{eq:genLLG_form}, this term vanishes due to the cross product with $\vec{m}$. Therefore, all terms quadratic in the rate of change of the magnetic moment, $\dotvec{m}$, vanish in the LLG equation.

To include all terms that are second order in the rate of change of $\vec{m}$, we must also consider terms that are linear in the second derivative of the temporal variation, i.e., the acceleration-like terms. The allowed terms are limited to
\begin{align}
    \vec{H}_\text{eff}^{(2)} = \eta_1 \ddotvec{m} + \eta_2 \vec{m} \times \ddotvec{m} + \eta_3 \vec{m} \times (\vec{m} \times \ddotvec{m}).
\end{align}
As in the first-order case, the $\eta_3$ term renormalizes the $\eta_1$ term. 
However, in contrast to the first-order case, the $\eta_2$ term must be retained. When inserted into Eq.~\eqref{eq:genLLG_form}, it yields
\begin{align}
    \eta_2 \vec{m} \times (\vec{m} \times \ddotvec{m}) = \eta_2 \left( \vec{m} (\vec{m} \cdot \ddotvec{m}) - \ddotvec{m} \right).
\end{align}
In the first-order case, we used the orthogonality condition $\vec{m} \cdot \dotvec{m} = 0$, but no such relation exists for the second derivative. Therefore, the $\eta_2$ term contributes nontrivially and is retained in the phenomenological description.
\subsection{General form}
Collecting all symmetry-allowed terms up to second order in time derivatives, the most general phenomenological form of the inertial Landau-Lifshitz-Gilbert (ILLG) equation for an isotropic system is (renaming  $\eta_2 \rightarrow \eta_1 \alpha_\eta$, $\gamma_1 \rightarrow - \gamma$ and removing subscripts)
\begin{align} \label{eq:llg_genform}
    \dotvec{m} = \vec{m} \times \left(
        -\gamma \vec{H} 
        + \alpha \dotvec{m} 
        + \eta \ddotvec{m} 
        + \eta \alpha_\eta \vec{m} \times \ddotvec{m} 
    \right).
\end{align}
Here, the $\gamma$-term governs the precessional dynamics, and $\alpha$ represents Gilbert damping. $\eta$ and $\alpha_\eta$ capture reactive and dissipative inertial effects. 
$\alpha$ and $\alpha_\eta$ are dimensionless.
$\alpha$ and $\alpha_\eta$ are independent coefficients of distinct symmetry-allowed structures; $\alpha_\eta$ is not the next term in a frequency expansion of $\alpha$.
The final expression of Eq.~\eqref{eq:llg_genform} respects the constraint $|\vec{m}(t)| = 1$ and includes all allowed contributions up to second order in the temporal variations.
¤
\section{The linearized inertial LLG equation}
We linearize the inertial LLG equation to determine how the system responds to harmonic drives, thereby extracting the resonance frequencies and linewidths of the dynamical modes.

Phenomenologically, the LLG equation can include precession, damping, inertia, and dissipative nutational terms:
\begin{align}
\label{eq:llg_sec3}
    \dotvec m = -\gamma \mu_0 \vec m \times \vec H + \alpha \vec m \times \dotvec    m 
    + \eta \vec m \times \ddot{\vec m} + \eta \alpha_\eta \vec m \times (\vec m \times \ddot{\vec m}),
\end{align}
as shown in Sec. \ref{sec:sm_phen}. The gyromagnetic ratio of Eq.~\eqref{eq:llg_sec3} is positive, $\gamma = g |e| / 2 m_e$.
 
We apply a strong static field along with a weak transverse drive. The magnetic field is given by
\begin{align}\label{eq:H_drive}
        \vec H = 
    \begin{pmatrix}
        \epsilon h_x(t)  \\ \epsilon h_y(t)  \\ \varsigma H_z
    \end{pmatrix}
    ,
\end{align}
where $H_z >0$, and $\epsilon \ll 1$.
The static field can be aligned or anti-aligned with the laboratory $ z$-axis, denoted by $\varsigma = \pm 1$. 
For these scenarios, we linearize using $\vec m(t) \simeq \varsigma \vec z + \delta \vec m^\varsigma$, and include only first-order terms in the small deviations from the equilibrium magnetization.

First, we note that the normalization of the magnetization is
\begin{align}
    1 = \vec m^2 = 1  + \varsigma 2 \delta m_z^\varsigma +[\delta \vec m^\varsigma]^2, 
\end{align}
which implies that 
\begin{align}
\label{eq:norm_delta_mz}
    \delta m_z^\varsigma = - \varsigma \frac12 [\delta \vec m^\varsigma]^2.
\end{align}
We expand the deviation of the unit vector along the magnetization from the $\pm \vec z$ direction in a parallel part and a perpendicular part:
\begin{align}
    \delta \vec m^\varsigma =  \mathbf z \delta m_z^\varsigma + \delta \vec m_\perp^\varsigma.
\end{align}
We expand in a small parameter $\epsilon$:
\begin{subequations}
    \begin{align}
    \delta m_z^\varsigma = \epsilon \delta m_z^{\varsigma,(1)} + \epsilon^2 \delta m_z^{\varsigma,(2)} + \mathcal O(\epsilon^3),\\
    \delta \vec m_\perp^\varsigma = \epsilon \delta \vec m_\perp^{\varsigma,(1)} + \epsilon^2 \delta \vec m_\perp^{\varsigma,(2)} + \mathcal O(\epsilon^3).
\end{align}
\end{subequations}
Eq.~\eqref{eq:norm_delta_mz} implies, by matching orders of $\epsilon$ to first and second order:
\begin{subequations}
\begin{align}
    \delta m_z^{\varsigma,(1)} &= 0,\\
    \delta m_z^{\varsigma,(2)}  &= - \varsigma \frac12 \bigl[\delta \vec m_\perp^{\varsigma,(1)}\bigr]^2.
\end{align}
\end{subequations}
Hence, to first order in $\epsilon$, only the transverse parts $\delta \vec m_\perp^{\varsigma, (1)}$ will contribute.
We denote the first-order corrections without the (1) superscript (labeling first-order corrections) from this point on. 
Inserting the parametrization of Eq.~\eqref{eq:H_drive} into Eq.~\eqref{eq:llg_sec3} and collecting all first‑order terms yield the coupled equations
\begin{subequations}
\label{eq:dotm}
\begin{align}
\label{dotmx}
    & \eta \alpha_\eta \delta \ddot m_x^\varsigma + \varsigma \eta \delta \ddot m_y^\varsigma + \delta \dot m_x^\varsigma + \varsigma \alpha \delta \dot m_y^\varsigma + \varsigma \omega_H \delta m_y^\varsigma =  \varsigma \gamma \mu_0  h_y,
    \\  \label{dotmy}
    & \eta \alpha_\eta \delta \ddot m_y^\varsigma - \varsigma \eta \delta \ddot m_x^\varsigma + \delta \dot m_y^\varsigma - \varsigma \alpha \delta \dot m_x^\varsigma - \varsigma \omega_H \delta m_x^\varsigma = - \varsigma \gamma \mu_0 h_x ,
\end{align}
\end{subequations}
where we introduced $\omega_H = \gamma \mu_0 H_z$ corresponding to the FMR precession frequency in the inertia-free case without dissipation.
We have $\omega_H >0$.
The corresponding equations for Eq. \eqref{eq:dotm} constitute two coupled linear differential equations, and are decoupled by transforming to a circular basis, defined by 
\begin{align} \label{eq:m_circbasis}
    \delta m_\pm^\varsigma = \delta m_x^\varsigma \pm \iu \delta m_y^\varsigma.
\end{align}
The linearized LLG equation in the circular basis is
\begin{align}
\label{eq:llg_circbas}
    \left[\eta ( \varsigma \pm \iu \alpha_\eta) \partial_t^2 + (\varsigma \alpha \pm \iu) \partial_t + \varsigma \omega_H \right] \delta m_\pm^\varsigma = \varsigma \gamma \mu_0 h_\pm ,
\end{align}
where we introduced the circular components of the driving field $h_\pm = h_x \pm \iu h_y $.
\subsection{Susceptibility}
We Fourier transform Eq. \eqref{eq:llg_circbas} using 
\begin{subequations} \label{eq:circ_FT}
\begin{align}
\delta  m_\pm^\varsigma(t) = \int_{-\infty}^\infty \frac{\d\omega}{2 \pi}  \delta m_\pm^\varsigma(\omega) \e^{\iu \omega t},
\\
h_\pm(t) = \int_{-\infty}^\infty \frac{\d\omega}{2 \pi} h_\pm(\omega) \e^{\iu \omega t},
\end{align}
\end{subequations}
and define the susceptibilities as 
\begin{align}
     \vec m_\pm^\varsigma(\omega) = \chi_\pm^\varsigma(\omega) h_\pm(\omega).
\end{align}
The susceptibilities are
\begin{align} \label{eq:chipm}
    \chi_\pm^\varsigma( \omega) = \frac{ \gamma \mu_0}{- \omega^2 \eta ( 1 \pm \iu \varsigma \alpha_\eta) + \iu \omega ( \alpha \pm  \iu \varsigma) +  \omega_H},
\end{align}
and satisfy the reality condition
\begin{align}
    [\chi_\pm^\varsigma(-\omega)]^* = \chi_\mp^\varsigma(\omega),
\end{align}
as well as the external field reversal symmetry relation
\begin{align}
    \chi_\pm^+(\omega) = \chi_\pm^-(\omega).
\end{align}
We obtain the poles of the susceptibilities by solving for the roots of the denominator of \eqref{eq:chipm}.  
We assume that the low-frequency precessional modes $\omega^\text{prec}$ and the high-frequency nutational modes $\omega^\text{nut}$ are well separated, and solve for the precessional and nutational modes by disregarding the $\omega^2$ terms and the $\omega^0$ terms, respectively:
\begin{subequations}
\begin{align}
    \omega^{\text{prec},\varsigma}_\pm &= \pm \varsigma \frac{\omega_H}{1 + \alpha^2} + \iu \alpha \frac{\omega_H}{1 + \alpha^2},
    \\
    \omega^{\text{nut},\varsigma}_\pm &= \mp \varsigma ( 1 - \alpha \alpha_\eta) \frac{\eta^{-1}}{1 + \alpha_\eta^2} + \iu (\alpha + \alpha_\eta  ) \frac{\eta^{-1}}{1 + \alpha_\eta^2}.
\end{align}
\end{subequations}
Now, since $\alpha \alpha_\eta \ll 1$, $\chi_+^+$ ($\chi_-^+$) has two resonance frequencies: one positive (negative) low-frequency mode dominated by $H_z$ and $\alpha$, and one positive (negative) high-frequency mode dominated by $\eta$ and the sum $\alpha + \alpha_\eta$.
The corresponding expressions for $\chi_+^-$ and $\chi_-^-$ carry the opposite signs.
In the main text, we restrict the results to the case where the external static field is parallel to the $z$-axis, i.e., $\varsigma = 1$, and disregard the $\varsigma$ superscript.
\section{Phenomenological expressions for the pumped current}
Phenomenologically, inertial terms can have both bulk and interfacial (spin‑pumping) contributions, analogous to Gilbert damping in a ferromagnet adjacent to a normal metal, where $\alpha = \alpha_\text{b} + \alpha_\text{sp}$. 
Thus, the pumped current in an adjacent NM should take the general form 
\begin{align}
    \vec j_s = a \vec m \times \dotvec    m 
    +b \vec m \times (\vec m \times \dotvec m) 
    +
    c \vec m \times \ddot{\vec m} + d \vec m \times (\vec m \times \ddot{\vec m}).
\end{align}
Here, we will expand the magnetization around its equilibrium value $\vec m(t) \simeq \varsigma \vec z + \delta \vec m^\varsigma$. 
We consider the phenomenological form of the spin current by expanding into different orders of the small quantity $\delta \vec m^\varsigma$. We allow for an equilibrium magnetization direction that is either parallel ($\varsigma = 1$) or antiparallel ($\varsigma = -1$) to the lab-frame coordinate $\vec z$, anticipating that reversing the external field will be a useful experimental probe.
We focus our attention on the dissipative contributions to spin-pumping, $\vec j_s^\text{dis}$, terms that are odd under time-reversal.  
We disregard the terms proportional to $b$ and $c$ since they do not constitute a dissipative dc spin current.

A similar expansion as in Sec. II shows that the $z$ component of the conventional spin pumping and nutational spin pumping to the lowest order in $\epsilon$ are
\begin{subequations} \label{eq:Jss}
    \begin{align}
    j_{s, z}^{\varsigma,\text{sp}} &= \epsilon^2 a \,
[\delta  m_x^{\varsigma} \delta \dot m_y^{\varsigma} - \delta \dot m_x^{\varsigma} \delta m_y^{\varsigma}],
\\
j_{s, z}^{\varsigma,\text{nsp}} &= \varsigma \epsilon^2 d \, [\delta m_x^{\varsigma} \delta \ddot m_x^{\varsigma} + \delta m_y^{\varsigma} \delta \ddot m_y^{\varsigma}].
\end{align}
\end{subequations}
From this point, we set $\epsilon = 1$, remembering that the result is valid to second order in the magnetic deviations from equilibrium.
In the following, we calculate the pumped spin currents, assuming a harmonic driving field, in the circular basis where the susceptibility is diagonal. 
\subsection{Harmonic driving field}
We consider a harmonic driving field:
\begin{subequations}
\label{eq:hxyt}
\begin{align}
    h_x(t) &= A_x \cos({\Omega t}), \\
    h_y(t) &= A_y \cos({\Omega t + \phi_h}),
\end{align}
\end{subequations}
where $A_x$ and $A_y$ are time-independent real and positive amplitudes, and $\phi_h\in(-\pi, \pi]$ is a phase that controls the polarization. Without loss of generality, we assume $\Omega \geq 0$ ($\Omega < 0$ corresponds to the opposite chirality, which is controlled by $\phi_h$).
The driving field of Eq.~\eqref{eq:hxyt} has Fourier components given by
\begin{subequations}
\begin{align}
    h_x(\omega) &= \frac{A_x}{2} 2 \pi [ \delta(\omega - \Omega) + \delta(\omega + \Omega)], \\
     h_y(\omega) &= \frac{A_y}{2} 2 \pi [ \delta(\omega - \Omega) \e^{\iu \phi_h} + \delta(\omega + \Omega) \e^{-\iu \phi_h}],
\end{align}
\end{subequations}
which leads to the circular fields
\begin{align}
    h_\pm(\omega) = h_x(\omega) \pm \iu h_y(\omega) = \pi \left[ A_\pm \delta(\omega - \Omega) + A_\mp^* \delta(\omega + \Omega) \right],
\end{align}
where $A_\pm = A_x \pm \iu \e^{\iu \phi_h} A_y$.
We proceed by transforming Eq.~\eqref{eq:Jss} into the circular basis defined in Eq.~\eqref{eq:m_circbasis}, and then use the Fourier transform from Eq.~\eqref{eq:circ_FT} as well as the diagonal susceptibilities in this basis, Eq.~\eqref{eq:chipm}, to arrive at the following form of the conventional and nutational spin pumping:
\begin{subequations} \label{eq:Jss_h_chi}
    \begin{align}
    \label{eq:Jss_h_chiP_prec}
    j_{s, z}^{\text{sp},+} &=  \frac{a}{4} \Omega \,
\left(|\chi_+^+(\Omega)|^2 |A_+|^2 - |\chi_-^+(\Omega)|^2 |A_-|^2 \right),
\\
j_{s, z}^{\text{sp},-} &=  \frac{a}{4} \Omega \,
\left(|\chi_+^-(\Omega)|^2 |A_+|^2 - |\chi_-^-(\Omega)|^2 |A_-|^2 \right),
\\
\label{eq:Jss_h_chi_nut}
j_{s, z}^{\text{nsp},+} &= -  \frac{d}{4} \Omega^2 \, \left(|\chi_+^+(\Omega)|^2 |A_+|^2 + |\chi_-^+(\Omega)|^2 |A_-|^2 + A_+ A_- \chi_+^+(\Omega) \chi_-^+(\Omega) \e^{2 \iu \Omega t} + A_+^* A_-^* \chi_+^{+*}(\Omega) \chi_-^{+*}(\Omega) \e^{-2 \iu \Omega t} \right),
\\
\label{eq:Jss_h_chi_nut_d}
j_{s, z}^{\text{nsp},-} &=   \frac{d}{4} \Omega^2 \, \left(|\chi_+^-(\Omega)|^2 |A_+|^2 + |\chi_-^-(\Omega)|^2 |A_-|^2 + A_+ A_- \chi_+^-(\Omega) \chi_-^-(\Omega) \e^{2 \iu \Omega t} + A_+^* A_-^* \chi_+^{-*}(\Omega) \chi_-^{-*}(\Omega) \e^{-2 \iu \Omega t} \right).
\end{align}
\end{subequations}
The first two terms in the nutational spin pumping of Eq.~\eqref{eq:Jss_h_chi_nut} and Eq.~\eqref{eq:Jss_h_chi_nut_d} are dc contributions, whereas the last two terms oscillate at frequency $2 \Omega$ and therefore represent ac components of the pumped spin current.
The ac contributions are, however, always suppressed relative to the dc terms because the product $\chi_+ \chi_-$ is small, since at least one of the susceptibilities is off resonance.
In the main text, we retain only the dc terms, which can be large at resonance.
The poles of $\chi_+^\varsigma$ occur at $\Omega \simeq \varsigma \omega_H$ and $\Omega \simeq - \varsigma \eta^{-1}$, while $\chi_-^\varsigma$ has poles at $\Omega \simeq -\varsigma \omega_H$ and $\Omega \simeq  \varsigma \eta^{-1}$.
In the main text, we restrict the results to the case where the external static field is parallel to the $z$-axis, i.e., $\varsigma = 1$, and disregard the $\varsigma$ superscript.
\subsection{Isolating the nutational spin current via FMR measurements}
Experimentally, the conventional and nutational spin currents in Eq.~\eqref{eq:Jss_h_chi} are convoluted, since they transform similarly under reversal of the external field and under changing the chirality of the driving field.
In this section, we show that measuring the differential spin current with respect to the external field, $\d / \d\omega_H j_{s, z}$, provides a clear way to separate the signal from the nutationally pumped current.
We assume a right-handed circularly polarized external drive, such that $A_- = 0$, and consider the frequency region close to the precession resonance frequency, disregarding the effect of nutation in this regime. 
This approximation is valid when the precession and nutation frequencies are well separated.
We consider an external field along the laboratory $z$ axis ($\varsigma = +1$) and approximate the Lorentzian distributions arising from the susceptibilities as delta functions close to the precession frequency $\omega_H$, valid when $\alpha \ll 1$:
\begin{align}
    |\chi_+^+(\Omega)|^2 \simeq \frac{(\gamma \mu_0)^2}{(\Omega - \omega_H)^2 + \alpha^2 \Omega^2} \rightarrow \frac{\pi (\gamma \mu_0)^2}{\alpha \omega_H} \delta(\Omega - \omega_H).
\end{align}
We integrate the spin current signals of Eqs.~\eqref{eq:Jss_h_chiP_prec} and \eqref{eq:Jss_h_chi_nut} over the FMR resonance, using integration limits wider than the linewidth.
We find that the integrated conventional spin pumping $X^\text{sp}_\text{FMR}$ is independent of the external field $\omega_H$, while the integrated nutational spin pumping $X^\text{nsp}_\text{FMR}$ is linear in $\omega_H$:
\begin{subequations}
\label{eq:jsp_int_FMR}
\begin{align}
    X^\text{sp}_\text{FMR} = \int \d\Omega \, j_{s, z}^{\text{sp},+} &\simeq \frac{a \pi (\gamma \mu_0)^2 |A_+|^2}{4\alpha }, \\
    X^\text{nsp}_\text{FMR} = \int \d\Omega \, j_{s, z}^{\text{nsp},+} &\simeq -\frac{d \pi (\gamma \mu_0)^2 |A_+|^2}{4\alpha } \omega_H.
\end{align}
\end{subequations}
A straightforward way to measure the nutational spin current is to calculate the differential integrated spin current,
\begin{align}
     \frac{\d}{\d \omega_H}  X_\text{FMR} =
    \frac{\d}{\d \omega_H} [X^\text{sp}_\text{FMR} + X^\text{nsp}_\text{FMR} ]  
    \simeq -\frac{d \pi (\gamma \mu_0)^2 |A_+|^2}{4\alpha }.
\end{align}
This signal is independent of the conventional (linear) pumped spin current and yields a clear, measurable effect.
\subsection{Isolating the nutational spin current via FMR and NR measurements}
Another approach to isolating the conventional and nutational spin pumping is first to repeat the procedure above at the NR frequency $\omega_\eta = \eta^{-1}$, using a left-handed field $A_-$:
\begin{subequations}
\label{eq:jsp_int_NR}
\begin{align}
    X^\text{sp}_\text{NR} = \int \d\Omega \, j_{s, z}^{\text{sp},+} &\simeq -\frac{a \pi (\gamma \mu_0)^2 |A_-|^2}{4 (\alpha + \alpha_\eta) }, \\
    X^\text{nsp}_\text{NR} = \int \d\Omega \, j_{s, z}^{\text{nsp},+} &\simeq -\frac{d \pi (\gamma \mu_0)^2 |A_-|^2}{4 (\alpha + \alpha_\eta) } \omega_\eta.
\end{align}
\end{subequations}
We add the results obtained from using a right-handed field at FMR (Eq.~\eqref{eq:jsp_int_FMR}) and a left-handed field at NR (Eq.\eqref{eq:jsp_int_NR}), and obtain the sum
\begin{align}
\label{eq:sum_X}
    X_\text{NR}+ X_\text{FMR } = 
    \frac{a \pi \gamma^2 \mu_0^2}{4} \left[ \frac{|A_+|^2}{\alpha} - \frac{|A_-|^2}{\alpha + \alpha_\eta}     \right]
    -
        \frac{d \pi \gamma^2 \mu_0^2}{4} \left[ \frac{|A_+|^2}{\alpha} \omega_H + \frac{|A_-|^2}{\alpha + \alpha_\eta} \omega_\eta     \right].
\end{align}
The constants $\alpha, \alpha_\eta, \omega_\eta$ and $\omega_H$ are known or can be measured independently. 
Thus, by measuring the sum of Eq.~\eqref{eq:sum_X} at the FMR frequency and NR frequency (using drives of opposite chirality), one can determine the two constants $a$ and $d$, fully characterizing the conventional and nutational spin pumping.

In summary, we have derived the general form of the pumped spin current in an NM adjacent to an FM.
The spin current along the $z$-direction contains both dc and ac components for the dissipative inertial term $\vec m \times (\vec m \times \ddotvec{m})$, in addition to the dc term arising from the standard spin pumping term $\vec m \times \dotvec{m}$. 
The dc contribution from the nutational spin current is isolated by measuring the slope of the spin current as $\omega_H$ is varied experimentally, or by comparing spin current measurements at the FMR and NR frequency.
\section{Scattering theory for a nonmagnetic metal-ferromagnet heterostructure}
We calculate the pumped spin current from a magnetic system by explicitly solving for the time-dependent scattering matrix using a time-dependent unitary rotation.
We consider a semi-infinite one-dimensional ferromagnet (FM) at $x>0$ in contact with a nonmagnetic metal (NM) at $x<0$. Electrons are incident from the left and reflected by the FM interface. We assume the FM is insulating, so there are no propagating modes in the FM.

The Hamiltonian of the system is
\begin{align} \label{eq:H_lab}
    H(x, \boldsymbol \rho, t) = H_0(x, \boldsymbol \rho) + \Theta(x) \left[ V(x) + \Delta \vec m(t) \cdot \bm{\sigma} \right].
\end{align}
Here, 
\begin{equation}
    H_0(x, \boldsymbol \rho) = -\frac{\hbar^2}{2m} \partial_x^2 + H_\perp(\boldsymbol \rho)
\end{equation}
represents the nonmagnetic Hamiltonian for a free particle in the $x$-direction, with transverse coordinate $\boldsymbol \rho$, which will result in transverse waveguide modes labeled by the index $n$. Moreover, $V$ is a charge potential and $\Delta$ is the exchange energy magnitude. The direction of the exchange field is a time-dependent unit vector $\vec m(t) = \begin{pmatrix}
\sin \theta \cos \phi,& \sin \theta \sin \phi,& \cos \theta\end{pmatrix}$, where $\theta$ and $\phi$ are the polar and azimuthal angles. The charge potential and exchange energy are assumed to be constant and time-independent, and active only inside the FM. 
The Schrödinger equation reads,
\begin{align} \label{eq:SE_lab}
    \iu \hbar \partial_t \psi(x, \vrho, t) =  H(x, \boldsymbol \rho, t) \psi(x, \vrho, t),
\end{align}
where the time-dependence of the wavefunction $ \psi(x, \vrho, t)$ is nontrivial due to the time-dependence of the Hamiltonian. 
\subsection{Transformation into rotating frame}
We solve the time‑dependent scattering problem by transforming to a frame that rotates uniformly about the $z$-axis.
To separate the rotating frame from the lab frame (where the Hamiltonian and wavefunctions are found as in Eq. \eqref{eq:H_lab} and \eqref{eq:SE_lab}), we denote quantities in the rotating frame of reference with a tilde $(\sim)$.
In the following, we consider a system undergoing simple and uniform precession: i.e., we assume that $\theta$ is a constant and $\phi = \omega t$ where $\omega$ is the angular frequency of the precessing ferromagnet. 
The rotation is performed according to the unitary transformation $ \psi(x, \vrho, t) = U(t) \tilde  \psi(x, \vrho, t)$, where $U$ is a time-dependent and unitary matrix:
\begin{align} \label{eq:Ut}
    U(t) = \e^{- \iu \omega t \sigma_z / 2 }.
\end{align}
This transformation greatly simplifies the ferromagnetic term,
\begin{align}
    U^\dag(t) H(x,\vrho, t) U(t) 
    &= H_0(x, \vrho) + \Theta(x) \left[ V(x) + \Delta \tilde{\vec m}_0 \cdot \bm{\sigma} \right],
\end{align}
because the exchange field is independent of the azimuthal angle $\phi$ in this frame of reference, $\tilde{\vec m}_0 = \begin{pmatrix}
        \sin \theta& 0 & \cos \theta.
    \end{pmatrix}$.
The rotation of the reference frame comes at the cost of introducing a gauge term due to the time-dependence in the transformation:
$
    \iu \hbar U^\dag(t) \partial_t U(t) = \hbar \omega \sigma_z / 2.
$
In the rotated frame, the total Hamiltonian reads
\begin{align}\label{eq:Hrot}
    \tilde H(x, \vrho) = 
    U^\dag(t)  H(x, \vrho, t) U(t) - \iu U^\dag(t) \partial_t U(t) 
    &= \tilde H_0(x, \vrho) + \Theta(x)  \left[ V(x) + \Delta \tilde{\vec m}_0 \cdot \bm{\sigma} \right] ,
\end{align}
and the Schrödinger equation is 
\begin{align}
    \iu \hbar \partial_t \tilde \psi^\text{N}(x, \vrho, t) =  \tilde H(x) \tilde \psi^\text{N}(x, \vrho, t). 
\end{align}
Here, we defined $\tilde H_0 = H_0 - \hbar \omega  \sigma_z / 2$ as the normal metal Hamiltonian in the rotating frame.
Crucially, the rotated (total) Hamiltonian $\tilde H$ in Eq. \eqref{eq:Hrot}, which includes the extra gauge term due to the time-dependent rotation, is time-independent.
We can thus solve for the eigenfunctions in the NM:
\begin{align}
    \tilde H_0(x, \vrho) \tilde \psi_{snE}^\text{N} = E \tilde \psi_{snE}^\text{N},
\end{align}
where $E$ is the single-particle energy, consisting of a transverse part $\epsilon_n$ and a longitudinal part that depends on the wave number and electron spin. 
A similar consideration shows that the eigenfunctions in the FM are $\tilde \psi_{\alpha n E}^\text{F}$, where $\alpha = \pm$ denotes the two spins, which are not parallel to the $z$-direction due to the presence of the exchange term.
These eigenfunctions allow us to solve this time-independent scattering problem analytically in this frame of reference. 
\subsection{The field operator}
We derive the field operator to find later the spin currents pumped by the precessing magnet. 
In the NM, in the rotating frame of reference, the Hamiltonian is spin-diagonal, and the solution to the Schrödinger equation with energy $E$, transverse mode $n$, and spin $s$ in the rotating frame is
\begin{align}\label{eq:phisE}
    \tilde \psi_{s n E}^\text{N}(x, \vrho, t) = \frac{1}{\sqrt{h v_s}} \e^{\pm \iu k_{sn} x}\e^{- \iu E t / \hbar} \phi_n(\vrho) \chi_s = \tilde \psi^\pm_{snE}(x, \vrho, t)  \chi_s,
\end{align}
where the spinors are $\chi_\uparrow =  (1, 0)$ and $\chi_\downarrow = (0, 1)$, we defined $\tilde \psi_{s nE}^{\pm}(x, \vrho, t) = \frac{1}{\sqrt{h v_{sn}}} \e^{\pm \iu k_{sn} x}\e^{- \iu E t / \hbar} \phi_n(\vrho) $, $\phi_n(\vrho)$ is the transverse part of the wavefunction with mode $n$, the spin-dependent wave vector $k_{sn}$ is defined through 
\begin{align} \label{eq:ks}
    \frac{\hbar^2}{2m} k_{sn}^2 - s \frac{\hbar \omega}{2}  + \epsilon_n = E,
\end{align}
and the velocity is defined as $v_{sn} = \hbar k_{sn} / m$. In Eq. \eqref{eq:ks}, $s = \pm 1$ for spin up/down.
In the definition of the scattering state in Eq. \eqref{eq:phisE}, we have chosen the prefactor to ensure $E$-normalization:
$
    \braket{\tilde \psi_{snE}^\text{N} | \tilde \psi_{s'm E'}^\text{N}} =\delta_{ss'} \delta_{nm} \delta(E - E').
$
Before proceeding, we discuss the energy and $k$-values in the lab frame and the rotating frame. In both frames, an incoming particle can be reflected with or without a spin flip. In the lab frame, the $k$-value of the incoming and reflected components is equal in magnitude, but the energy of the reflected particle with a spin flip will differ from the incoming energy with a sideband energy $\pm \hbar \omega$.
In the rotating frame, the energy is conserved, but the $k$-value is different in the reflected electron with a spin flip compared to the incoming and reflected electron without a spin flip.

We consider an incident electron (from the left) with spin $s$ and transverse mode $n$, which is reflected with spin- and mode-dependent reflection coefficients in the rotating frame. 
The scattering state with energy $E$ for an incoming electron in the NM reads
\begin{align}
\label{eq:phi_upE}
    \tilde \psi_{s n E}^N(x, \vrho, t) = \tilde \psi_{s n E}^{+ } \chi_s 
    +    \sum_{\sigma m} \tilde \psi_{\sigma m E}^{- }  \chi_\sigma \tilde r_{\sigma s}^{mn},
\end{align}
where $\tilde r_{s \sigma}^{mn}$ denotes the components of the (time-independent) reflection matrix in the lab frame. 
We now promote the single-particle scattering states to second-quantized field operators, which allows us to compute expectation values of spin currents
In the lab frame, the field operator is found by applying the unitary transformation $U(t)$ to the rotating frame field operator:
\begin{align}
    \hat{\Psi}(x, t) = U(t) \hat{\tilde \Psi}(x, t) =  \sum_{s  \sigma nm}  \int \d E   \hat{\tilde a}_{s n}(E) U(t) \chi_\sigma  \left[\delta_{s \sigma} \delta_{nm} \tilde \phi_{sE}^+(x, \vrho, t)   + \tilde r_{ \sigma s}^{mn} \tilde \phi_{\sigma E}^-(x, \vrho, t) \right],
\end{align}
where $\hat{\tilde a}_{sn}(E)$ annihilates an electron with spin $s$, transverse mode $n$, and energy $E$ in the NM in the rotating frame. From the definition of $U(t)$ in Eq. \eqref{eq:Ut}, we see that the effect is to produce an extra spin-dependent exponential term $U(t) \chi_\sigma = e^{-\iu \sigma  \omega t / 2}$. 
Since $k_{\sigma m}(E) = k_m(E + \sigma \hbar \omega / 2 )$ and similar for the velocity $v_{\sigma m}$,  we can rewrite the field operator in terms of energy-shifted quantities by introducing $\tilde E = E + \sigma \hbar \omega /2$. 
We introduce $\psi_{n \tilde E}^\pm(x, \vrho, t) = \frac{1}{\sqrt{h v_n(\tilde E)}} \e^{\pm \iu k_n(\tilde E) x} \e^{- \iu \tilde E t / \hbar} \phi_n(\vrho) $, and write:
\begin{align}
        \hat{\Psi}(x, \vrho, t) = \sum_{s \sigma nm} \int \d \tilde E  \hat{\tilde a}_{sn}(\tilde E - \sigma \hbar \omega / 2) \chi_\sigma \left[\delta_{s \sigma}  \delta_{nm}\psi_{n \tilde E}^+(x, \vrho, t)  \chi_s
        + \tilde r_{ \sigma s}^{mn}(\tilde E - \frac{\sigma \hbar \omega}{2})  \psi_{\tilde E}^-(x, \vrho, t) \right].
\end{align}
Finally, we replace the rotating frame annihilation operator $\hat a$ with the lab frame annihilation operator $\hat{\tilde a}$ via the relation $\hat{\tilde a}_{s n}(\tilde E - s \hbar \omega /2) = \hat a_{s n}(\tilde E)$, and rename the integration variable for simpler notation.
The field operator in the lab frame is thus given by
\begin{align} \label{eq:Psi}
    \hat{\Psi}(x, \vrho, t) = \sum_{sn} \int \d E \biggl[\psi_{n E}^+(x, \vrho, t)  \chi_s \hat a_{sn}(E) + \sum_{\sigma m} \psi_{m E}^-(x, \vrho, t) \chi_\sigma \hat a_{sn} \bigl(E - \frac{\sigma - s}{2} \hbar \omega \bigr) \tilde r_{\sigma s}^{mn}\bigl(\omega, E - \frac{\sigma \hbar \omega}{2}\bigr) \biggr] ,
\end{align}
where we write the $\omega$-dependence of the reflection matrix explicitly. Thus, we have managed to find the field operator written in terms of the \textit{lab frame} annihilation operators. The annihilation operators in the lab frame are easy to work with, since we assume that the system is in equilibrium in the lab frame, so that the electrons are distributed according to the Fermi-Dirac function. The reflection matrix $\tilde r$ is, however, defined in the rotating frame, where we can find analytical solutions. 
\subsection{The spin current to second order in precession frequency}
Using the field operator from Eq. \eqref{eq:Psi}, we can now determine the $z$-component of the expectation value of the spin current in the lab frame:
\begin{align} \label{eq:jz_gen}
    J_{s,z} =  \frac{\hbar^2}{2 m }  \int \d\vrho \Im{\langle \hat{ \Psi}^\dag(x, \vrho, t) \sigma_z \partial_x \hat{  \Psi}(x, \vrho, t) \rangle }.
\end{align}
The spin current has a unit of angular momentum per unit time, or equivalently, energy.
We find the expectation value of the spin current by inserting the field operator from Eq. \eqref{eq:Psi} in the spin current expression in Eq.~\eqref{eq:jz_gen}:
\begin{align}
    J_{s, z}  =
    \frac{1}{4 \pi} \int \d E  \sum_{s \sigma nm} \sigma f(E - \frac{\sigma - s}{2}\hbar \omega ) \left[ \delta_{\sigma s} \delta_{nm} - |r_{\sigma s}^{mn}(\omega, E - \frac{\sigma \hbar \omega}{2})|^2 \right].
\end{align}
where we used $\langle \hat a_{sn}^\dag(E - \frac{\sigma - s}{2} \hbar \omega) \hat a_{s'm}(E' - \frac{\sigma - s}{2} \hbar \omega')
\rangle = \delta_{ss'} \delta_{nm} \delta(E - E') f(E - \frac{\sigma - s}{2} \hbar \omega)$ in the lead in the lab frame, and where we have dropped the tilde on the reflection matrix for simpler notation.
From this point on, the reflection matrix is always given in the rotating frame, where it is time-independent and where we can obtain analytical expressions.
Next, we shift the energy integral by $\sigma\hbar \omega/2$ to find a simpler expression:
\begin{align} \label{eq:jz}
    J_{s, z} =
    \frac{1}{4 \pi} \int \d E  \sum_{s \sigma nm} \sigma f(E  + \frac{s\hbar \omega}{2} ) \left[ \delta_{\sigma s} \delta_{nm} - | r_{\sigma s}^{mn}(\omega, E)|^2 \right] .
\end{align}
Eq.~\eqref{eq:jz} will serve as the starting point for finding the spin current up to second order in the precession frequency $\omega$.
We expand the Fermi-Dirac function
\begin{align}
    f(E + \frac{ s\hbar \omega}{2} ) \approx f(E) +  \frac{s\hbar \omega}{2} f'(E) + \frac{1}{2} \frac{\hbar^2 \omega^2}{4} f''(E),
\end{align}
and the reflection matrix (disregarding the spin, transverse modes, and $E$ arguments for notational simplicity):
\begin{align}
    r(\omega) \approx r + \omega r' + \frac{1}{2} \omega^2  r'',
\end{align}
where $r = r(\omega = 0)$ and $r' = \partial_\omega r(\omega)|_{\omega = 0}$.
These relations, in turn, imply that we can write
\begin{align}
    \bigl|r(\omega)|^2 \approx |r|^2 + 2 \omega \Re{r^*r'} + \omega^2\bigl(\Re{r^* r''} + |r'|^2 \bigr),
\end{align}
for all matrix elements of the reflection matrix.
To simplify the expressions we obtain in the following, we utilize that the reflection matrix is unitary (in our system, there are no current-carrying modes in the FM) to all orders of $\omega$, i.e., $r r^\dag = r^\dag r = 1$. 
For the reflection matrix components, this implies that $\sum_{\sigma m} |r_{\sigma s}^{mn}(\omega)|^2 = \sum_{s n} |r_{\sigma s}^{mn}(\omega)|^2 = 1$. To zeroth, first, and second order in $\omega$,  we find that unitarity implies
\begin{subequations}
\begin{align}
    \sum_{\sigma m} |r_{\sigma s}^{mn}|^2 = \sum_{s n} |r_{\sigma s}^{mn}|^2 = 1 \label{eq:unit0}
    \\
     \sum_{\sigma m} \Re{r_{\sigma s}^* r_{\sigma s}'  } = \sum_{s n} \Re{r_{\sigma s}^* r_{\sigma s}'  } =  0 \label{eq:unit1} 
     \\
    \sum_{\sigma m} \bigl(\Re{r_{\sigma s} ^* r_{\sigma s}''} +|r_{\sigma s}'|^2  \bigr) 
    = 
    \sum_{s n} \bigl(\Re{r_{\sigma s} ^* r_{\sigma s}''} +|r_{\sigma s}'|^2  \bigr) = 0. \label{eq:unit2}
\end{align}
\end{subequations}
To zeroth order, the contribution to the spin current is
\begin{align}
    J_{s, z}^{(0)} = -\frac{1}{4 \pi} \int \d E  f(E) \sum_{\sigma nm  } \sigma \sum_s  |r_{\sigma s}|^2 = 0,
\end{align}
where we used unitarity of the reflection matrix, Eq. \eqref{eq:unit0}.

The first-order contribution to the spin current is
\begin{align} \label{eq:j1}
    J_{s, z}^{(1)} = -\frac{\hbar \omega}{4 \pi} \sum_{nm} \left( |\rud^{mn}| + |\rdu^{mn}|^2 \right)_F
    - \frac{\omega}{2\pi} \int \d E f(E) \sum_{s \sigma} \sigma   \Re{r_{\sigma s}^*r_{\sigma s}'},
\end{align}
where $F$ denotes evaluation at the Fermi energy $E_F$.
Again, we use the unitarity of the reflection matrix, Eq. \eqref{eq:unit1}, to see that the second term in Eq. \eqref{eq:j1} vanishes.
In our (rotated) frame of reference, the zeroth order in $\omega$ reflection matrix is $r = r_+ + r_- \vec m_0 \cdot \vec \sigma$, where $r_\pm = \frac12 (r_\uparrow \pm r_\downarrow)$, and $r_{\uparrow(\downarrow)}$ describes the diagonal reflection matrix in the reference frame where the magnetization is pointing along the $z$-direction.
This is because the quantization axis in the NM is along the $\tilde{\vec m_0}$ direction to zeroth order in $\omega$.
By introducing the mixing conductance $g_{\uparrow \downarrow} =  \sum_{mn} (\delta_{mn} - r_\uparrow^{mn} (r_\downarrow^{mn})^* )$, we can cast the first term in terms of the real part of the mixing conductance:
\begin{align} \label{eq:jz1}
    J_{s, z}^{(1)} = -\frac{\hbar \omega}{4 \pi} g^r_{\uparrow \downarrow} \sin^2 \theta.
\end{align}
The imaginary part of the mixing conductance does not appear in the spin current $J_{s, z}$ in the case of simple precession.

To second order, we find three contributions to the spin current: one from expanding the distribution function to second order, one from expanding the reflection matrix to second order, and one from the cross term: 
\begin{subequations}
\begin{align}
    J_{s, z}^{(2),A}  &= -\frac{\hbar^2 \omega^2}{16 \pi} \sum_{nm}\int \d E  f''(E) \left(|\rud^{mn}|^2 - |\rdu^{mn}|^2 \right) = 0 
    \\
     J_{s, z}^{(2),B} &= \frac{\hbar \omega^2}{4 \pi} \sum_{s \sigma nm} s \sigma \Re{(r_{\sigma s}^{mn})^* (r_{\sigma s}^{mn})'}_F
     \\
      J_{s, z}^{(2),C} &= -\frac{\omega^2}{4 \pi} \sum_{nm}\int \d E  f(E) \sum_{s \sigma} \sigma \left[\Re{(r_{\sigma s}^{mn})'' (r_{\sigma s}^{mn})^*} + |r_{\sigma s}'|^2  \right] = 0,
\end{align}
\end{subequations}
where we again used unitarity, Eq. \eqref{eq:unit2},  to cancel the first and third contributions.
The remaining contribution to the second-order spin current is
\begin{align}
\label{eq:jz2}
    J_{s, z}^{(2)} = \frac{\hbar \omega^2}{4 \pi} \sum_{s \sigma nm} s \sigma \Re{(r_{\sigma s}^{mn})^* (r_{\sigma s}^{mn})'}_F , 
\end{align}
and represents corrections to the frozen scattering matrix given by derivatives of the reflection matrix, evaluated at the Fermi level. 
In the following, we explicitly calculate the reflection matrix to first order in $\omega$ and insert the result to obtain the second-order nutational spin current pumped. This expansion provides a systematic approach to compute spin-dependent reflection and spin pumping to quadratic order in the precession frequency.
\subsection{Solving the scattering problem to first order in reflection matrix}
The Hamiltonian of the NM-FM system is defined in the rotating frame as
\begin{equation}
H = -\frac{\hbar^2}{2m}\partial_x^2 
+ \Theta(x) \left[ V + \Delta\, \tilde{\vec m}_0\cdot\bm{\sigma} \right]
- \frac{\hbar\omega}{2} \sigma_z.
\label{eq:H_full}
\end{equation}
In the FM, we write this as 
\begin{equation}
H_\text{FM} = -\frac{\hbar^2}{2m}\partial_x^2 
+ \vec d \cdot \boldsymbol \sigma 
\label{eq:H_FM}
\end{equation}
where we introduced $\vec d = \begin{pmatrix}
    \Delta \sin \theta, & 0, & \Delta \cos \theta - \hbar \omega / 2
\end{pmatrix} = |\vec d| \vec n_0$, where $\vec n_0$ is a unit vector that lies in the $xz$ plane. It will be convenient in the following to write this in terms of a polar angle $\beta$ as
$\vec n_0 = \begin{pmatrix}
    \sin \beta, & 0, & \cos \beta
\end{pmatrix}$.
Assuming the FM supports only evanescent modes at the energy of interest ($E < V \pm |\vec d|$), the general solution in the FM region ($x>0$) is
\begin{equation}
\phi_{\rm FM}(x) = A_+ \, \chi_+ \, e^{-\kappa_+ x} + A_- \, \chi_- \, e^{-\kappa_- x},
\end{equation}
where $\chi_\pm$ are the spin eigenspinors of $\vec d \cdot\bm{\sigma}$ with eigenvalues $\pm 1$, given by 
\begin{align}
    \chi_+ = \begin{pmatrix}
        \cos{(\beta / 2)} \\ \sin{(\beta / 2)}
    \end{pmatrix}
    && 
    \chi_- = \begin{pmatrix}
        - \sin{(\beta / 2)} \\ \cos{(\beta / 2)} \\
    \end{pmatrix},
\end{align}
and
\begin{equation}
\kappa_\pm = \sqrt{\frac{2m[(V + E_\pm) - E]}{\hbar^2}},
\end{equation}
where we defined the part of the energy coming from the spin as
\begin{align}
    E_\pm = \pm \sqrt{\Delta^2 - \Delta \hbar \omega \cos \theta + \frac{\hbar^2 \omega^2}{4} }.
\end{align}
The NM region ($x<0$) supports propagating plane waves,
\begin{align}
    \phi_\text{NM} =  \e^{\iu k_s x} \chi_s 
    +    \sum_\sigma \e^{-\iu k_\sigma x}  \chi_\sigma r_{\sigma s},
\end{align}
where $k_s = \sqrt{2m( E + s \hbar \omega / 2)}/ \hbar$.
Continuity of the wavefunction and its derivative at $x = 0$ gives
\begin{subequations}
\begin{align}
\label{eq:bdc_wf}
    \chi_s + \sum_\sigma r_{\sigma s} \chi_\sigma &= A_{+s} \chi_+ + A_{-s} \chi_-
    \\
    \label{eq:bdc_dwf}
    \iu k_s \chi_s - \iu \sum_\sigma k_\sigma \chi_\sigma r_{\sigma s} &= - K_+ A_{+s} \chi_+ -  K_- A_{-s} \chi_-.
\end{align}
\end{subequations}
For a given $s$, this represents four linear equations that we can solve to obtain the four unknown coefficients.
\subsubsection{Matrix form}
To solve equations \eqref{eq:bdc_wf} and \eqref{eq:bdc_dwf}
analytically, we introduce the matrices
\begin{align}
    \bar X_\text{N} = (\chi_\uparrow, \chi_\downarrow) 
    && \bar X_\text{F} = (\chi_+, \chi_-) 
    && \bar k = \text{diag}(k_\uparrow, k_\downarrow)
    &&
    \bar K = \text{diag}(K_+, K_-) && \bar A = \begin{pmatrix}
        A_{+ \uparrow} & A_{+ \downarrow}\\
        A_{- \uparrow} & A_{- \downarrow}
    \end{pmatrix},
\end{align}
and denote the $2\times 2$ reflection matrix as $\bar R$. The $\bar \ldots$ denote $2\times 2$ matrices.
In this notation, the boundary conditions read
\begin{subequations}
\begin{align}
\label{eq:bdc_wf_mat}
    \bar X_\text{N} + \bar X_\text{N} \bar R &= \bar X_\text{F} \bar A
    \\
    \label{eq:bdc_dwf_mat}
    \iu \bar X_\text{N} \bar k - \iu \bar X_\text{N} \bar k \bar R &= -\bar X_\text{F} \bar K \bar A.
\end{align}
\end{subequations}
We can, e.g., isolate the unknown coefficient matrix $\bar A$ using the first equation, and insert it into the second equation to find
\begin{align} \label{eq:sol_R}
    R = (\iu \bar k - \bar M)^{-1} (\iu \bar k + \bar M),
\end{align}
where we introduced $\bar M = \bar X_F \bar K \bar X_F^{-1}$, and used that $\bar X_N = \bar 1$ to simplify the result.
In the following, we will expand to first order in $\omega$: $R = R^{(0)} + \omega R^{(1)}$.
\subsubsection{Zeroth-order scattering matrix}
To lowest order, we find $E_\pm = \pm \Delta$ (assuming $\Delta >0$), so that $K_\pm \rightarrow \kappa_\pm = \sqrt{2m[(V \pm \Delta) - E]}/\hbar$. Moreover, $k_s \rightarrow k = \sqrt{2mE}/\hbar $ and $\beta \rightarrow \theta$.
Imposing continuity of the wavefunction and its derivative at $x=0$, we reproduce the conventional form of the zeroth-order scattering matrix for a magnet pointing in the $\tilde{\vec m_0}$ direction:
\begin{equation}
R^{(0)} = r^{(0)}_0 \sigma_0 + r^{(0)}_m (\tilde{\vec m}_0 \cdot \boldsymbol \sigma),
\end{equation}
where
\begin{align}
    r_0^{(0)} =\frac{(k^2 + \kappa_+\kappa_-)}{(k + \iu \kappa_+) (k + \iu \kappa_-)}
    &&
    r_m^{(0)} = \frac{- i k (\kappa_+ - \kappa_-)}{(k + \iu \kappa_+) (k + \iu \kappa_-)}.
\end{align}
Similarly, we can expand the reflection matrix in Eq.\eqref{eq:sol_R} to any order.
\subsubsection{Spin current}
Finally, we insert the reflection matrices to zeroth and first order into the spin current expressions given in Eqs.~\eqref{eq:jz1} and \eqref{eq:jz2}.
We define dimensionless quantities $\tilde V = V / E_F, \tilde \Delta = \Delta / E_F$, and $ \tilde K_\pm = K_\pm / k_F$, where $k_F = \sqrt{2m E_F} / \hbar$ is the Fermi momentum.
We can rewrite the spin current expressions to matrix form:
\begin{subequations}
\begin{align}
    J_{s, z}^{(1)} = - \frac{\hbar \omega}{8 \pi} \tr{1 - \sigma_z (R^{(0)})^\dag \sigma_z R^{(0)}}
    \\
    J_{s, z}^{(2)} =  \frac{\hbar \omega^2}{4 \pi} \tr{ \sigma_z (R^{(0)})^\dag \sigma_z R^{(1)}}
\end{align}
\end{subequations}
The first-order spin current then becomes 
\begin{align}
    J_{s, z}^{(1)}= - \frac{\hbar \omega}{4 \pi} \sin^2 \theta \frac{2 (\kappa_+ - \kappa_-)^2}{(1 + \kappa_+^2) (1 + \kappa_-^2)},
\end{align}
and the second-order pumped current becomes
\begin{align}
    J_{s, z}^{(2)} = - \frac{\hbar^2 \omega^2}{4 \pi E_F} \sin^2 \theta \cos \theta \frac{2 (\kappa_+ - \kappa_-)^2 \left( \kappa_+ \kappa_-  - \frac12 \tilde \Delta \frac{\kappa_+ + \kappa_-}{\kappa_+ - \kappa_-}  \right)}{\tilde \Delta \kappa_+ \kappa_- ( 1 + \kappa_+^2) ( 1 + \kappa_-^2)}.
\end{align}
Dividing the second-order spin current by the first-order spin current, 
introducing the function
\begin{align}
    \Pi(\tilde \Delta, \tilde V) = \frac{\kappa_+ \kappa_-  - \frac12 \tilde \Delta \frac{\kappa_+ + \kappa_-}{\kappa_+ - \kappa_-}}{\tilde \Delta \kappa_+ \kappa_-},
\end{align}
and renaming $J_{s, z}^{(1)} \rightarrow J_{s, z}^\text{sp}$ and $J_{s, z}^{(2)} \rightarrow J_{s, z}^\text{nsp}$, we obtain Eq.~(12) in the main text.